\documentclass[%
 reprint,
superscriptaddress,
 amsmath,amssymb,
 aps,
]{revtex4-1}

\pdfoutput=1

\usepackage{soul}
\usepackage{graphicx}
\usepackage{dcolumn}
\usepackage{bm}

\usepackage{mathtools}

\usepackage[dvipsnames]{xcolor}

\usepackage[colorlinks=true,citecolor=blue]{hyperref}
\usepackage[normalem]{ulem}
\usepackage[caption=false]{subfig} 
\usepackage{amsmath}

\newcommand{\UIB}{Departament de F\'isica, Universitat de les Illes Balears, IAC3 -- IEEC, Crta. Valldemossa km 7.5, E-07122 Palma, Spain}
\newcommand{\APC}{Universit\'e de Paris, CNRS, Astroparticule et Cosmologie, F-75013 Paris, France}

\definecolor{granadagreen}{HTML}{078931}

\begin{document}


\title{IMRPhenomTP: A phenomenological time domain model for dominant quadrupole gravitational wave signal of coalescing binary black holes }

\author{H\'ector Estell\'es}
\affiliation{ \UIB}

\author{Antoni Ramos-Buades}
\affiliation{ \UIB}

\author{Sascha Husa}
\affiliation{ \UIB}

\author{Cecilio Garc\'ia-Quir\'os}
\affiliation{ \UIB}

\author{Marta Colleoni}
\affiliation{ \UIB}

\author{Le\"ila Haegel}
\affiliation{ \UIB}\affiliation{\APC}

\author{Rafel Jaume}
\affiliation{ \UIB}


\date{\today}

\begin{abstract}

In this work we present IMRPhenomTP, a time domain phenomenological model for the dominant $l=2$, $m=|2|$ modes of coalescing black hole binary systems and its extension to describe general precessing systems within the ``twisting up'' approximation. The underlying non-precessing model is calibrated to the new release of Numerical Relativity simulations of the SXS Collaboration and its accuracy is comparable to the state-of-the-art non-precessing dominant mode models as IMRPhenomX and SEOBNRv4. The precessing extension allows for flexibility choosing the Euler angles of the time-dependent rotation between the co-precessing and the inertial reference systems, including the single spin NNLO and the double spin MSA PN descriptions present in other models, numerical integration of the orbit averaged spin evolution equations, different choices for the evolution of the orbital angular momentum norm and a simple approximation to the ringdown behaviour. 
\end{abstract}

\pacs{Valid PACS appear here}
\maketitle


\section{\label{intro}Introduction}


During the last decade, a rich variety of waveform models that describe gravitational waves at all the stages of coalescence of a quasi-circular BBH system have been constructed, based on different strategies but all relying on the crucial information provided by numerical relativity (NR) simulations about the strong field regime of the binary merger. Currently, a variety of models are available for non-precessing systems, including models for the dominant quadrupole emission of non-precessing systems \cite{khan2016frequency}, \cite{pratten2020setting}, \cite{Boh__2017} and precessing systems \cite{Hannam_2014}, \cite{Khan_2019},  \cite{Nagar_2018}, for higher order multipole emission of non-precessing systems \cite{garcaquirs2020imrphenomxhm}, \cite{London_2018}, \cite{Cotesta_2018}, \cite{Varma_2019_2}, and for higher order multipole emission of precessing systems \cite{khan2019including}, \cite{aless2020multipolar}, \cite{Varma_2019}, \cite{phenomxphm}, \cite{SEOBNRv4PHM:inprep}, enabling to perform accurate parameter estimation of the source properties of gravitational waves events. As the sensitivity of the aLIGO \cite{aLIGO2015} and AdVirgo \cite{Acernese_2014} interferometers is improving, many more binary systems, exhibiting a larger parameter space of such objects, are expected to be detected in a near future. Further improving the generality and accuracy of waveform models is an ongoing effort of the community, in particular concerning general precessing systems and eccentric systems.

The goal of this paper is to extend the scope of the IMRPhenom waveform family \cite{Ajith_2007, Santamar_a_2010, khan2016frequency, Hannam_2014, London_2018, Khan_2019, khan2019including, pratten2020setting, garcaquirs2020imrphenomxhm}, which are commonly referred to as phenomenological waveform models. These models are built in terms of piecewise closed form expressions, and until now have been constructed in the Fourier domain, in order to achieve fast evaluation times in gravitational wave data analysis procedures, where frequency domain templates are needed to compute the noise-weighted inner product with detectors data. However, in order to develop strategies for modelling generic binary systems, insight can be gained from a time domain description of the signals, where dynamical information of the system can be approached in a more direct way. For this reason, in this work, we present the first steps of a time domain phenomenological modelling framework with the aim of providing complementary strategies that could benefit the overall program on modelling accurately general BBH systems. We present IMRPhenomTP, a time domain phenomenological model for the dominant quadrupole emission of precessing systems. The core model is a phenomenological description of the $l=2$, $m=2$ mode for non-precessing systems calibrated to a dataset of 531 non-precessing numerical relativity simulations from the last release of the SXS Collaboration catalogue, and validated against a dataset of EOB-NR hybrids constructed from the same dataset. The precessing extension is based on the common approach of ``twisting up'' the non-precessing core model developed for the frequency-domain IMRPhenomP model \cite{Hannam_2014}.

In section \ref{sec:modelling} we describe the modelling strategy for the dominant quadrupole $l=2$, $m=|2|$ mode of non-precessing systems, splitted in three domains: inspiral, merger and ringdown, as well as the construction of the full waveform. In section \ref{sec:calib} we describe the calibration procedure to NR simulations, and in section \ref{sec:validation} we validate the model against a catalogue of hybrid waveforms and we offer comparison with other models. In \ref{sec:precessing} we describe the procedure for extending the model to precessing systems and we discuss strategies for further improvements in the precessing description.

\section{Aligned-spin modelling}\label{sec:modelling}



Gravitational wave detectors measure a projection of the two wave polarisations, $h_+$ and $h_{\times}$. The polarisations carry information about the intrinsic properties of the binary system and information about the orientation of the system. For quasi-circular BBH systems, where orbit eccentricity is negligible, the masses $m_{1,2}$ and the individual spins $\boldsymbol{S}_{1,2}$ of each black hole describe the physical state of the system at each time and therefore the emitted gravitational radiation. The total mass of the system $M=m_1+m_2$ is a scaling parameter that can be factorized employing geometrized units where $G=c=1$. Therefore, only a subset of 7 parameters is needed to describe the evolution of a particular system and a common parameterization is in terms of the symmetric mass ratio $\eta=(m_1m_2)/M^2$ and the dimensionless individual spins $\boldsymbol{\chi}_{1,2}=\boldsymbol{S}_{1,2}/m^2_{1,2}$. For non-precessing systems, the individual spins are parallel to the orbital angular momentum of the binary and the direction $\hat{\boldsymbol{L}}$ and magnitude remains approximately constant during the evolution so it is sufficient to consider the norm of each dimensionless spin $\chi_{1,2}=\hat{\boldsymbol{L}}\cdot\boldsymbol{\chi}_{1,2}$. Therefore the space of intrinsic parameters needed to describe non-precessing BBH systems is:
\begin{equation}
    \boldsymbol{\lambda}=\{\eta,\chi_1,\chi_2\}.
\end{equation}
The information about the relative orientation of a particular source with respect to the observer can be parametrized by the inclination $\iota$ between the orbital angular momemtum direction and the line-of-sight of the source and the orbital azimuthal angle $\phi$ that specifies the orientation of the binary in the orbital plane at a particular time.

A common approach to model the polarisations is to decompose the complex combination in a spin-weighted spherical harmonic (SWSH) basis, where the orientation is encoded in the basis functions and the coefficients, the gravitational wave modes, encode the intrinsic physical information:
\begin{equation}
h(t,\boldsymbol{\lambda},\iota,\phi)\equiv h_{+}-ih_{\times}=\sum_{l,|m|\leq l}h_{lm}(t,\boldsymbol{\lambda})\ \,_{-2}Y_{l,m}(\iota,\phi)\ ,
\end{equation}
where the complex functions $h_{lm}$ can be related to the emission of different multipole moments of the stress-energy tensor of the source. In order to simplify the modeling procedure, it is useful to represent the modes in polar form and to model separately amplitude and phase:
\begin{equation}
h_{lm}(t)=H_{lm}(t)e^{i\psi_{lm}(t)}\ .
\end{equation}
With a suitable reference frame for the decomposition on the sphere, the dominant contribution of the radiation is given by the quadrupolar emission $h_{2,|2|}$. Therefore, we follow the usual procedure applied in waveform modelling and choose to model first the $l=2$, $m=2$ mode for aligned-spin configurations, obtaining the $l=2$, $m=-2$ by symmetry:
\begin{equation}
h_{2,-2}(t)=(-1)^l h^*_{22}(t).
\end{equation}

In the inspiral regime, where both black holes orbit each other in quasi-circular orbits, the frequency of the different SWSH modes can be related to the orbital frequency of the binary as $\omega_{lm}(t)\simeq m \dot{\phi}_{orb}(t)$ where this approximation holds to a good accuracy until the minimum energy circular orbit (MECO) frequency \cite{Cabero_2017}, as discussed in \cite{garcaquirs2020imrphenomxhm}. After the merger, the remnant black hole relaxes to a stable state through damped emission, commonly known as the ringdown, in which the wave frequency of each mode approaches a fixed frequency known as ringdown frequency: $\omega_{lm}(t>t_{peak})\rightarrow \omega^{RD}_{lm}$. In the non-perturbative regime of the plunge-merger of the two black holes, accurate information about the emission frequency of the radiation is only known through the results of full NR simulations.

In this work we present the phenomenological modelling of the amplitude and phase derivative of the $l=m=2$ mode: $H_{22}(t)$ and $\omega_{22}(t)\equiv \dot{\phi}_{22}(t)$. We split the waveform into three physically motivated regions: the inspiral region, where analytical approximations are known in the Post-Newtonian (PN) framework \cite{Blanchet_2014}, the plunge-merger region, where information from numerical relativity (NR) simulations is crucial, and the ringdown region, where a combination of NR information and perturbation theory is needed. We set the boundary between the inspiral and merger regions at the time where the minimum energy circular orbit (MECO) frequency is achieved, since a well-informed PN description should be valid up to this frequency \cite{Cabero_2017} as it has been shown succesfully in \cite{pratten2020setting}. The boundary between the merger and ringdown is set at the peak time of the amplitude $H_{22}(t)$, which we set without loss of generality at $t=0$.

\subsection{\label{sec:inspiral}Inspiral emission}

While the binary objects are sufficiently far apart from each other, the weak field and low velocity conditions are satisfied and the binary dynamics can be adressed with the PN framework \cite{Blanchet_2014}. While a complete solution of the PN equations of motion still requires numerical integration of the full system of equations, further simplifications can be obtained assuming the adiabatic emission condition which allows to set the balance equations (see e.g.~\cite{buonanno2009comparison}):
\begin{subequations}
\begin{equation}
\label{eq:balanceeq1}
\dfrac{d\phi}{dt}-\dfrac{v^3}{M}=0,
\end{equation}
\begin{equation}
\label{eq:balanceeq2}
\dfrac{dv}{dt}+\dfrac{\mathcal{F}(v)}{ME'(v)}=0,
\end{equation}
\end{subequations}
where $v=( -\omega_{22}/2)^{1/3}$, $\phi(t)$ is the orbital phase of the binary, $\mathcal{F}(v)$ is the gravitational wave luminosity and $E(v)$ is the binding energy of the system.

The TaylorT family of gravitational wave templates (see \cite{buonanno2009comparison} for a presentation and systematic comparison) consists of different PN expansions of the ratio $\mathcal{F}(v)/E'(v)$, starting with TaylorT1 that expands each quantity in the quotient and then solve numerically equation (\ref{eq:balanceeq2}) with a ratio of polynomials. An alternative procedure, known as TaylorT4 \cite{TaylorT4}, expands instead the full quotient. A further step yields the TaylorT2 approximant, integrating the ratio of polynomials to consistent PN order and then obtaining a pair of parametric equations for $\phi(v)$ and $t(v)$ that can be solved numerically. One then can invert the relation $t(v)$ to obtain an analytical and explicit expression for $\phi(v(t))$ as a function of time, obtaining the TaylorT3 approximant \cite{blanchet2002gravitational}:
\begin{subequations}
\begin{equation}
\label{eq:taylort3eq1}
\phi^{(T3)}_{n/2}(t)=\phi_{ref} + \phi_{N}(t)\sum_{k=0}^{n}\hat{\phi}_k\theta^k,
\end{equation}
\begin{equation}
\label{eq:taylort3eq2}
\omega^{(T3)}_{n/2}(t)=\omega_{N}(t)\sum_{k=0}^{n}\hat{\omega}_k\theta^k,
\end{equation}
\end{subequations}
where $\theta(t)=[\eta(t_{0}-t)/(5M)]^{-1/8}$, $\omega_{N}=\theta^3/8$ and the PN coefficients $\hat{\omega}_k$ are given in Appendix \ref{sec:appendixOmega}.

TaylorT3 has the advantage of providing a closed-form expression of the orbital phase and consequently the orbital frequency as a function of time. However, as shown in \cite{buonanno2009comparison}, the maximum frequency for which TaylorT3 can reproduce accurately the inspiral phase is in general lower than for the other approximants. In fact, TaylorT3 becomes singular at $t=t_0$, which corresponds to different frequencies depending on the intrinsic parameters. Moreover, $t_0$ changes with the PN order employed in the expansion, so the common phenomenological strategy of correcting the high frequency behaviour by adding extra pseudo-PN terms with unknown coefficients and calibrating those with NR is bad-suited in this case, because one needs to previously know the appropriate $t_0$, which is not possible before the extra coefficients have a value. The strategy we follow in this work is to set from the beginning $t_0=0$, corresponding to the merger time, for all cases and then correct both at high and low frequencies with extra coefficients calibrated through collocation points. In some sense it is justified since $t_0$ should correspond to the merger time, but the incomplete information in TaylorT3 makes that for some cases this time is underestimated or overestimated. Since we are improving the accuracy though extra higher order terms, seems reasonable to fix the merger time information to the actual merger time.

We found that  one needs to extend the currently known 3.5 PN TaylorT3 to at least 6 pseudo-PN order to achieve an accurate description of the frequency and the phase until at least the minimum energy circular orbit time ($t_{\text{MECO}}$), i.e, adding 5 extra pseudo-PN terms at the corresponding orders:
\begin{equation}
\omega^{\text{insp}}_{22}(t)=\omega^{(T3)}_{3.5}(t) + \omega_{N}(t)\sum_{k=8}^{12}\hat{c}_k\theta^k,
\end{equation}
where the 5 extra pseudo-PN terms $\hat{c}_k$ are obtained imposing that the frequency match 5 collocation points. We set 4 collocation points in the late inspiral region to extend the validity at high frequencies, in particular at $t=-2000M$, $t=-1000M$, $t=2 t_{\text{MECO}}$ and $t=t_{\text{MECO}}$, and one collocation point at lower frequencies for compensating the shift at early frequencies caused by imposing $t_0=0$, in particular at $t=-10^5M$.

\begin{figure}
    \includegraphics[width=\linewidth]{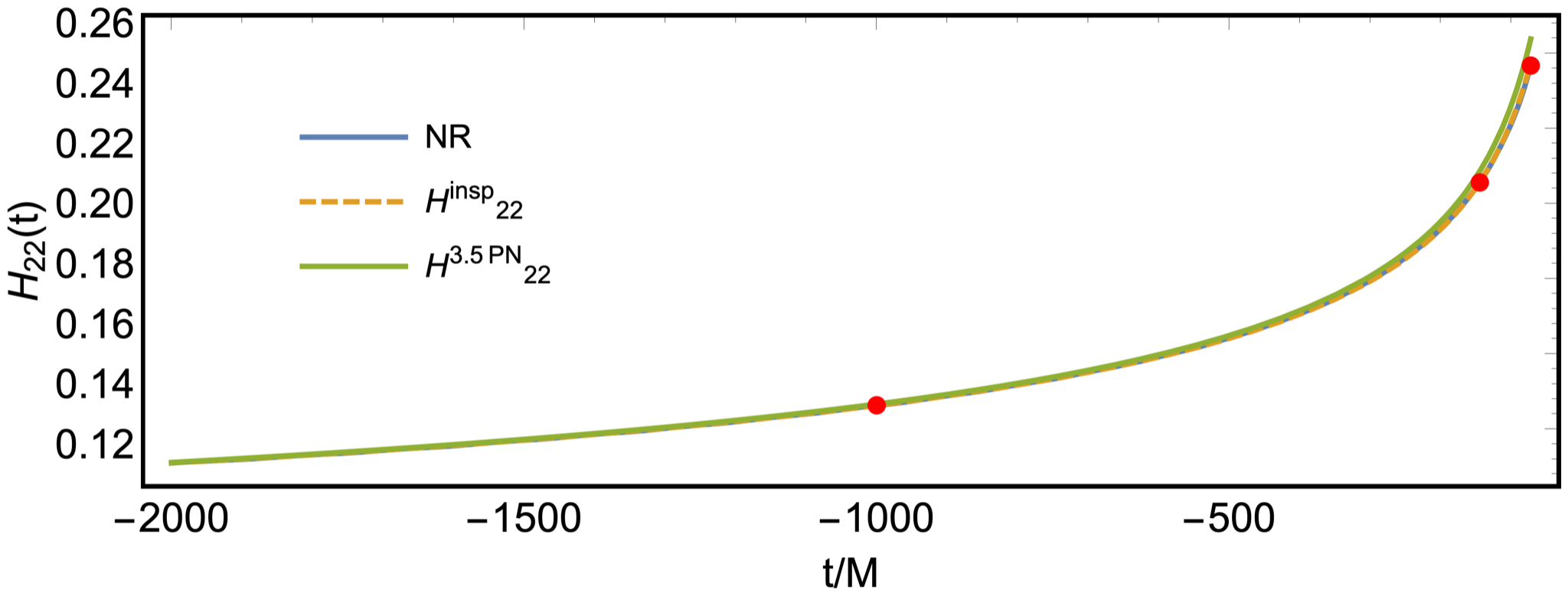}
    \includegraphics[width=\linewidth]{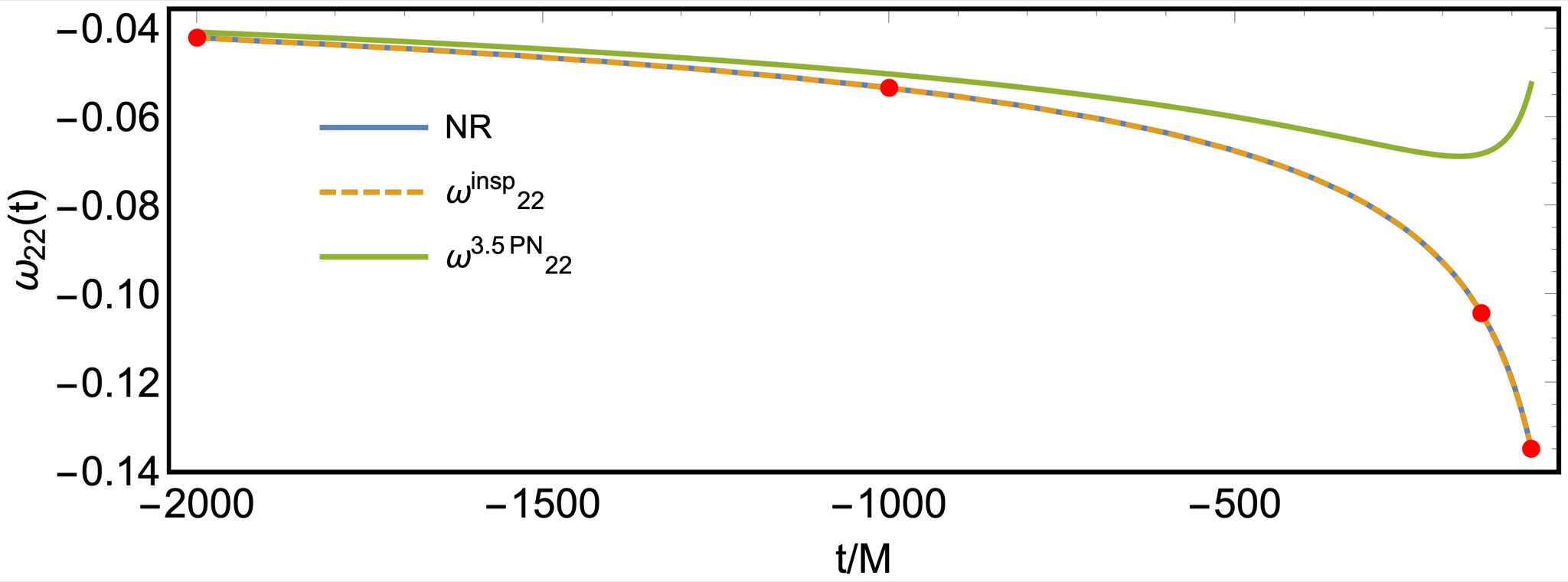}
    \includegraphics[width=\linewidth]{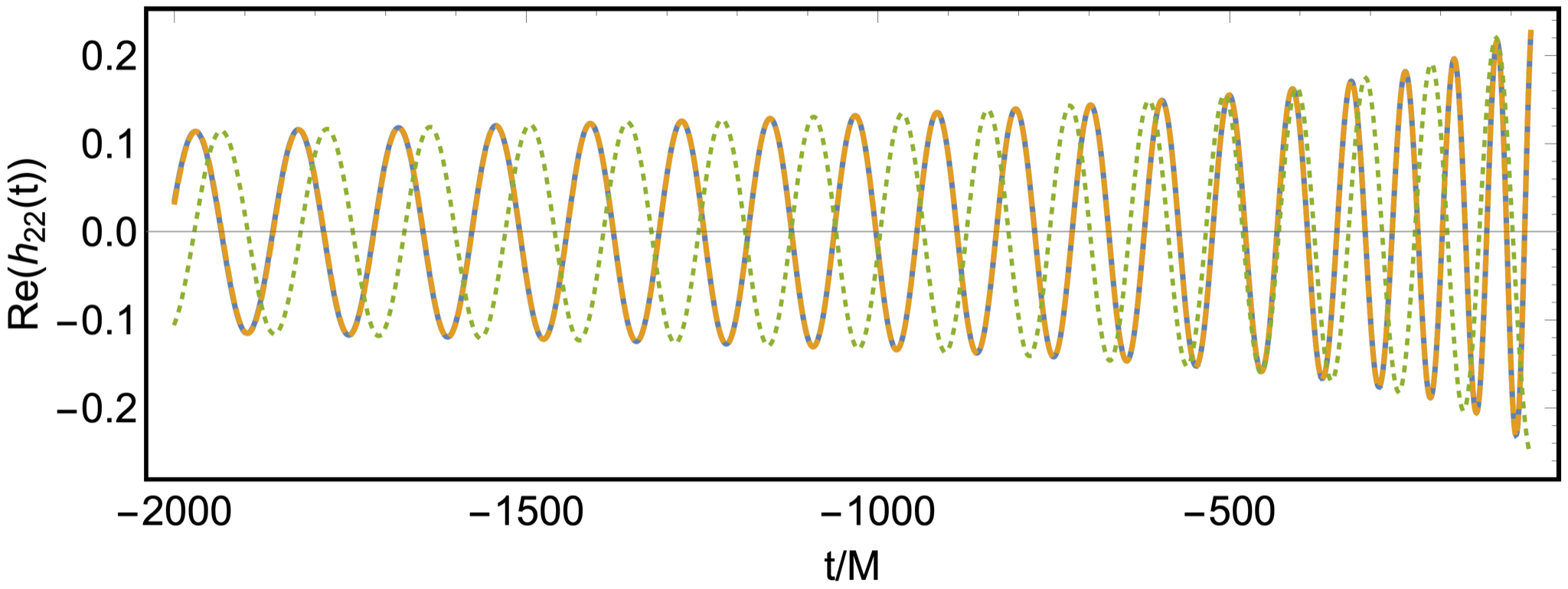}
  \caption{Late inspiral comparison with SXS:BBH:0001 NR simulation ($q=1$, $\chi_{1,2}=0$). Collocation points for amplitude and frequency are marked as red points.}
\end{figure}

During the inspiral, the amplitude of the emitted radiation can be computed with higher order PN extensions of the quadrupole formula. The 3PN expression for non-spinning BBH can be found in \cite{Blanchet_2008} and we also include the 3.5PN non-spinning corrections of \cite{faye2012postnewtonian}. In the spin sector, we include the 1.5PN spin contribution from \cite{arun2008} and the 2PN contribution from \cite{Buonanno_2013}:
\begin{equation}
\label{eq:pnamp}
    H_{22}^{3.5\text{PN}}(x)=2\eta\sqrt{\dfrac{16}{5}}x \sum_{k=0}^{7}\hat{h}_k x^{k/2}
\end{equation}
where $x(t)=(-\omega_{22}(t)/2)^{2/3}$ and the coefficients $\hat{h}_k$ are given in Appendix \ref{sec:appendixAmp}. However, this expression is not accurate enough to describe the emission close to the MECO because of the failure of the Post-Newtonian expansion as we approach the merger. Nevertheless, as in the case of the frequency, one can add higher order unknown PN terms to make the expression converge at least until the MECO time. We found that adding three extra terms, i.e. completing to 5 pseudo-PN order, the resulting expression works well across the parameter space:
\begin{equation}
H_{22}^{\text{insp}}(t)=H_{22}^{3.5\text{PN}}(x) +2\eta\sqrt{\dfrac{16}{5}}x \sum_{k=8}^{10}\hat{d}_k x^{k/2},
\end{equation}
where $\hat{d}_k$ are unknown coefficients determined by requiring the amplitude to match three collocation points at $t=-1000M$, $t=2 t_{\text{MECO}}$ and $t=t_{\text{MECO}}$.

\subsection{\label{sec:ringdown} Ringdown emission}

The damped emission of the final black hole can be well approximated by in terms of linear perturbations of the Kerr solution, for which analytical approximations are known in the framework of perturbation theory in terms of a linear combination of damped quasinormal modes (QNMs) \cite{Kokkotas_1999}:
\begin{equation}\label{eq:QNM_decomp}
h^{RD}_{lm}(t)=\sum_{n=1}^{\infty}c_{nlm}\exp[i\sigma_{nlm}(t-t_0)]
\end{equation}
where indices $n$, $l$ and $m$ refer to the energy level and to the spin-weighted spheroidal mode, $c_{lnm}$ are amplitude coefficients and $\sigma_{nlm}$ is the complex frequency of the mode level, from which the asymptotic final frequency $\omega_{nlm}^{\text{RD}}$ and damping frequency $\alpha_{nlm}$ can be obtained:
\begin{subequations}
\begin{align}
    \omega_{nlm}^{\text{RD}} &= \Re (\sigma_{nlm}),\\
    \alpha_{nlm} &= \Im (\sigma_{nlm}).
\end{align}
\end{subequations}

Here the frequencies $\omega_{nlm}^{\text{RD}}$ and $\alpha_{nlm}$ are known functions of the
spin of the Kerr black hole (see e.g.~\cite{Berti_2009}), which can be predicted from the initial masses and spins of the binary in terms of fits to numerical relativity data such as \cite{jimenez2017hierarchical}, which we will use in this work. The amplitudes and relative phases of the exponentially damped modes have to be computed with numerical relativity.
While the linear combination (\ref{eq:QNM_decomp}) is accurate in the late ringdown regime (but not so late that power-law tails take over), for the early ringdown modifications are required to accurately represent the signal. We follow a strategy first attempted in \cite{damour2014new}, which consists of factoring out the dominant QNM ground state $n=1$ from the waveform:
\begin{equation}
    \bar{h}_{lm}(t)=e^{i\sigma_{1lm}(t)}h_{lm}(t),
\end{equation}
and then proposing phenomenological ansatzs for the resulting amplitude and phase:
\begin{equation}
\label{eq:damouramp}
|\bar{h}_{22}(t)|=e^{\alpha_1(t)}|h(t)|=d_1\tanh[d_2 t+d_3] + d_4,
\end{equation}
\begin{equation}
\label{eq:NagarOmega}
\bar{\omega}_{22}(t)=\omega_{22}(t) - \omega^{RD}_{122}=c_1\dfrac{c_2(c_3e^{-c_2t} + 2c_4e^{-2c_2t})}{1+c_3e^{-c_2t} + c_4e^{-2c_2\ t}},
\end{equation}
where $\alpha_i\equiv \alpha_{i22}$, $d_1$ and $d_4$ fix the amplitude at the peak and ensure that the amplitude's derivative vanishes there, $c_2=d_2=(\alpha_2-\alpha_1)/2$, and $c_3$, $c_4$, $d_3$ are free phenomenological coefficients to fit to numerical data. To better understand the behavior of the frequency ansatz, it can be seen that with $c_4=0$ it is equivalent to a $\tanh$ function with a ``slope'' equal to the difference in damping frequencies between the first QNM overtone and the QNM ground state, analogous to the amplitude ansatz. We have checked that these ansatzs produce accurate results across the parameter space and are well suited for a future extension of the model with the inclusion of subdominant harmonics of the signal.

\begin{figure}[h]
  \centering
  \includegraphics[width=.95\linewidth]{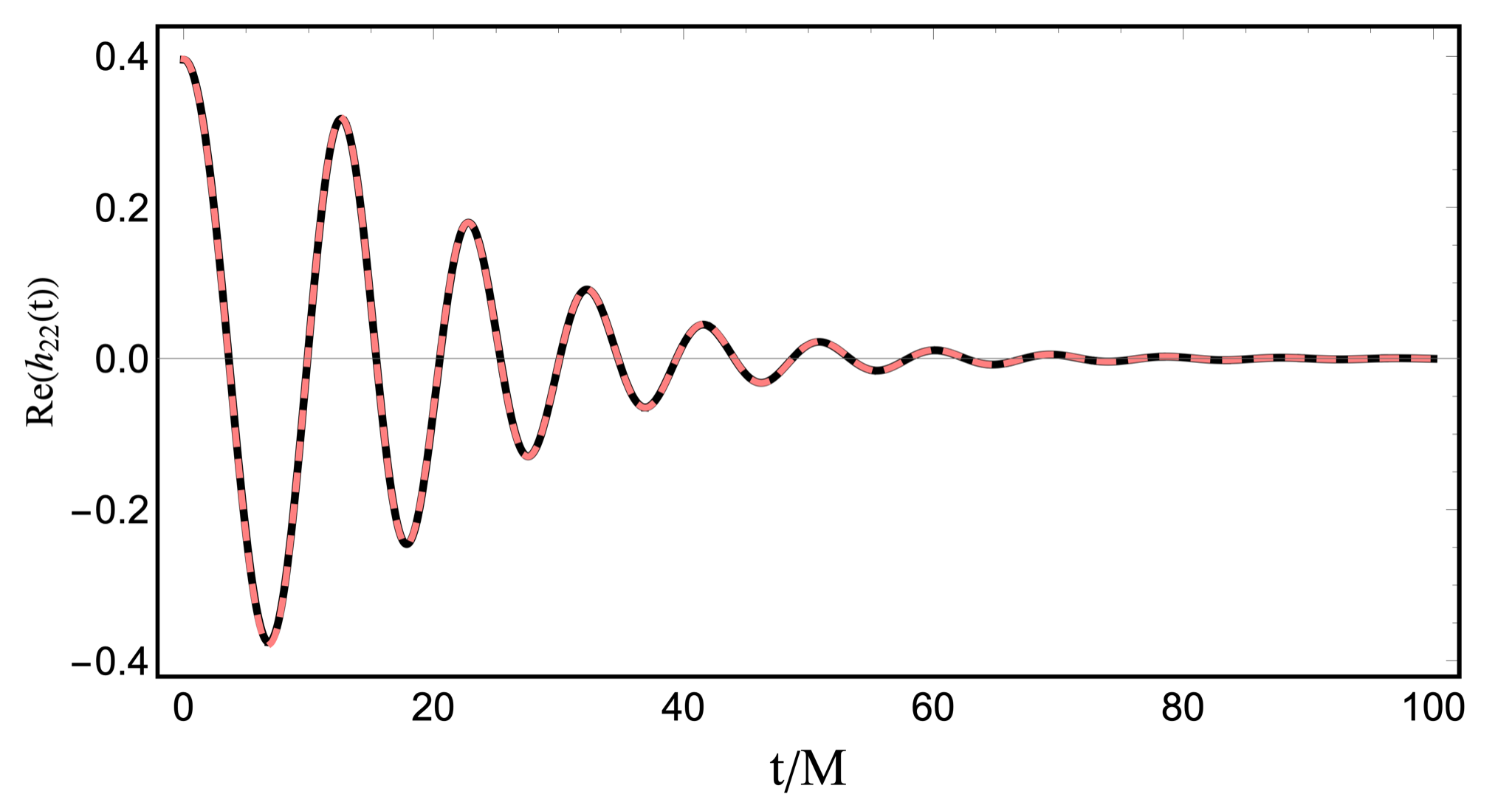}  
\caption{Comparison of ringdown ansatz with SXS:BBH:0152 with parameters $q=1$, $\chi_1=0.6$ and $\chi_2=0.6$.}
\label{fig:ringdown}
\end{figure}


\subsection{\label{sec:merger} Merger emission}
     
As the binary's frequency grows the PN description of the system looses accuracy, and eventually becomes invalid. Despite the complexity of the full nonlinear set of Einstein equations, numerical relativity simulations have shown that the evolution of binary systems retains a certain simplicity during the plunge-merger, before entering the ringdown regime described in Sec.~\ref{sec:ringdown}. In particular, for quasi-circular non-precessing systems, the orbital frequency remains monotonically increasing, accelerating as in the inspiral until the luminosity energy peak time and then decelerating to a constant final value of the relaxation frequency of the remnant black hole.


An interesting proposal to exploit the simplicity of the merger-ringdown phenomenology, focusing on the transition from the emission peak to the ``clean'' ringdown regime has been the study \cite{baker2008mergers} for non-spinning BBH, which has been extended to aligned-spin BBH in \cite{kelly2011mergers}. These works perform an analysis subtracting the asymptotic relaxation frequency to the waveform frequency and then proposing a hyperbolic tangent-like phenomenological ansatz to fit the growing behavior. For the amplitude, they rely on the implicit rotational source idea to connect the amplitude of the strain-rate (the time derivative of the strain) with the energy loss of the system through an effective moment of inertia that will correspond to a rigid rotating multipole that depends on the final state quantities. A more recent proposal to model the merger-ringdown strain amplitude is made in \citep{mcwilliams2019analytical}, where the author argues that from light-ring reflection considerations, the $\psi_4$ (second time derivative of the strain) amplitude around the peak is well modeled by a hyperbolic secant function that depends on the final state damping frequency. We have investigated these lines of research and found that while they can constitute a decent approximation, in particular the implicit rotational source model, it is difficult to connect in a smooth way the end of the inspiral with these descriptions since their accuracy before the strain peak is limited.

Motivated by the idea of employing the final state damping frequency and the hyperbolic function dependence of the previous studies, we propose the following phenomenological ans{\"a}tze for the frequency and the amplitude:
\begin{equation}
\label{eq:customomega}
\omega^{\text{merger}}_{22}(t)=\sum_{k=0}^{k=4}a_k\text{arcsinh}^k(\alpha_1t),
\end{equation} 
\begin{equation}
\label{eq:customamp}
H_{22}^{\text{merger}}(t)=b_0 + b_1 t^2 + b_2 \text{sech}^{1/7}(\alpha_1 t) + b_3 \text{sech}(\alpha_1 t),
\end{equation}
where in (\ref{eq:customomega}), for the frequency, $a_i$ are set requiring continuity and differentiability at the boundaries and imposing the frequency to match a collocation point at $t=0.25t_{\text{MECO}}$. In (\ref{eq:customamp}), for the amplitude, $b_i$ are also set requiring continuity and differentiability at the boundaries and demanding the amplitude to match a collocation point at $t=0.5t_{\text{MECO}}$. In both ansatzs, the damping frequency $\alpha_1$ of the groundstate QNM for the $l=2$, $m=2$ mode, presented in the previous section, is employed.

The linear combination of hyperbolic arcsin functions in the frequency allows the sufficient flexibility in modeling the growing rate of the frequency between the end of the inspiral and the ringdown, where different effects from high PN terms and final state non-perturbative effects produce a huge variability of the frequency growing rate. For the amplitude, the characteristic deformed bell-shape around the strain peak is well modelled by a combination of different powers of an hyperbolic secant with width given by the damping frequency of the final state. An additional advantage of these ansatz is that much of the freedom is set by the boundaries with the inspiral and ringdown regions, where physically motivated ansatzes are implemented. 

\begin{figure}[h]
  \centering
  \includegraphics[width=.95\linewidth]{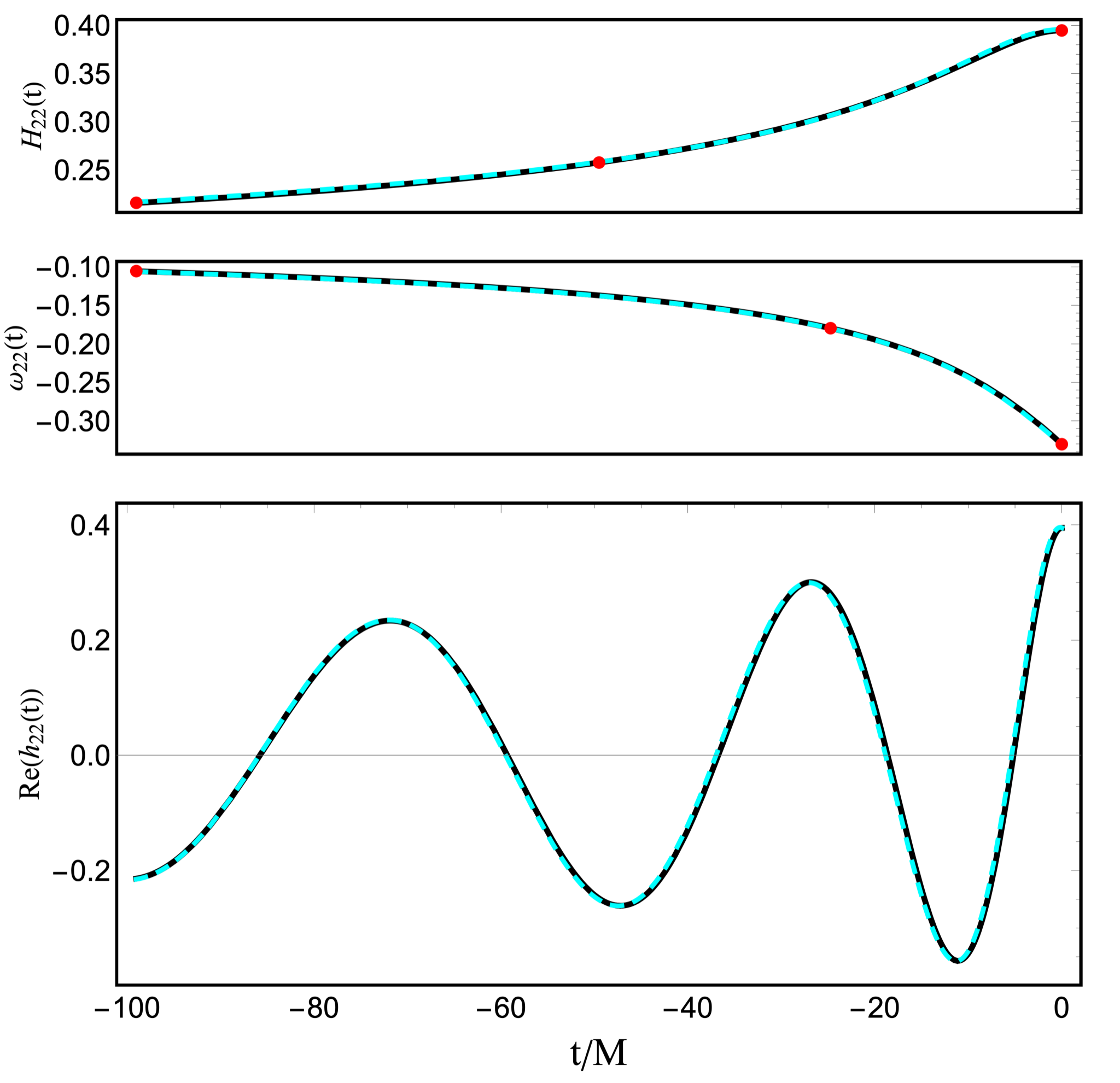}  
\caption{Comparison of merger amplitude and frequency ansatz with SXS:BBH:0210 NR simulation, with parameters $q=1$, $\chi_1=-0.9$ and $\chi_2=0$. Collocation points for amplitude and frequency are marked as red points.}
\label{fig:merger}
\end{figure}

\subsection{\label{sec:imr} Complete IMR Waveform}

Combining the ans{\"a}tze of the three regions from the previous sections and demanding continuity and differentiability at the interfaces of the regions, we obtain the complete $C^1$ inspiral-merger-ringdown (IMR) expressions for both amplitude and frequency: 

\begin{equation}
\label{eq:fullIMRomega}
  \omega_{22}(t) =
  \begin{cases}
                                   \omega^{\text{insp}}_{22}(t) & t\leq t_{\text{MECO}} \\
                                   
                                   \\\omega^{\text{merger}}_{22}(t)  & t_{\text{MECO}}\leq t \geq 0 \\
                                   \\
  								 \omega^{\text{RD}}_{22}(t) & t \geq 0,
  \end{cases}  
\end{equation}

\begin{equation}
\label{eq:fullIMRamp}
       H_{22}(t) =
 	 \begin{cases}
                                    H^{\text{insp}}_{22}(t) & t\leq t_{\text{MECO}} \\
                                   
                                   \\H^{\text{merger}}_{22}(t)  & t_{\text{MECO}}\leq t \geq 0 \\
                                   \\
  								 H^{\text{RD}}_{22}(t) & t \geq 0.
 	 \end{cases}
\end{equation}

In order to construct the IMR waveform, we need to integrate the frequency in time for obtaining the gravitational wave phase:
\begin{equation}
\phi_{22}(t)=\int\text{dt}\ \omega_{22}(t)\ .
\end{equation}
The expressions describing the inspiral, merger and ringdown regimes of the frequency can be integrated analytically, leading to closed form expressions for the IMR phase. The integration constants are set to guarantee  continuity between regions and to set a reference phase at a selected time.

\begin{figure}
    \includegraphics[width=1\linewidth]{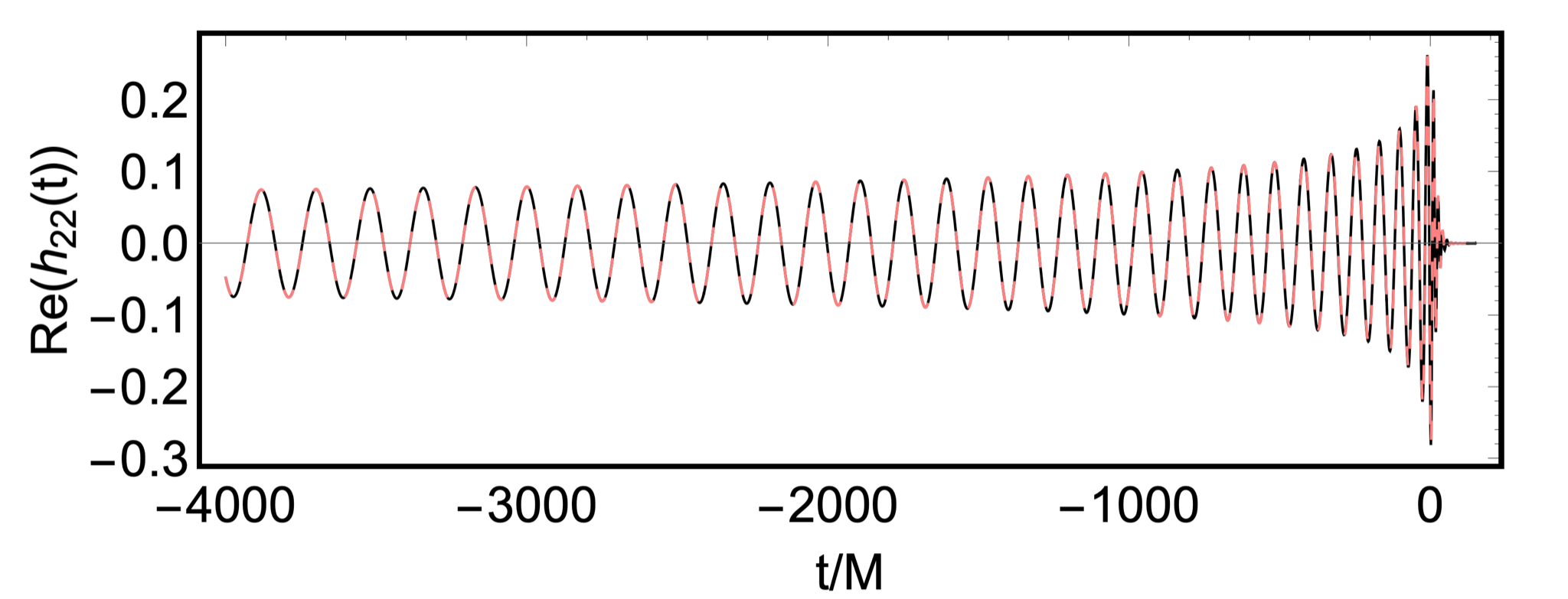}
    \label{fig:pnamp}
  \caption{Comparison of the complete IMR waveform with SXS NR simulation SXS:BBH:2139}
\end{figure}

\section{\label{sec:calib}Calibration}

The  complete IMR model constructed from (\ref{eq:fullIMRomega}) and (\ref{eq:fullIMRamp}) depends on a set of phenomenological coefficients that have to be calibrated to NR simulations across the three-dimensional parameter space $\{\eta,\chi_1,\chi_2\}$ that corresponds to non-precessing quasi-circular black hole binaries. In total, we have the following phenomenological coefficients:
\begin{itemize}
\item Inspiral frequency coefficients:
\begin{equation}
\hat{c}_8,\ \hat{c}_9,\ \hat{c}_{10},\ \hat{c}_{11},\ \hat{c}_{12}  	
\end{equation}
\item Inspiral amplitude coefficients:
\begin{equation}
\hat{d}_8,\ \hat{d}_{9},\ \hat{d}_{10} 	
\end{equation}
\item Merger frequency coefficients:
\begin{equation}
a_{0},\  a_{1},\  a_{2},\  a_{3},\  a_{4},\ a_{5},\    	
\end{equation}
\item Merger amplitude coefficients:
\begin{equation}
b_{0},\  b_{1},\  b_{2},\  b_{3},
\end{equation}
\item Ringdown frequency coefficients:
\begin{equation}
c_{1},\  c_{2},\  c_{3},\  c_{4}
\end{equation}
\item Ringdown Amplitude coefficients:
\begin{equation}
d_{1},\  d_{2},\ d_{3},\  d_{4}.
\end{equation}

\end{itemize}

As discussed in \cite{khan2016frequency, pratten2020setting} it is often advantageous to re-parameterize phenomenological coefficients in terms of collocation points, in particular to improve the conditioning of the calibration procedure. Taking into account that the collocation points placed at the boundaries between regions can be employed for solving coefficients in both regions, the following set of 16 quantities has to be calibrated:
\begin{itemize}
\item Amplitude collocation points:
\begin{equation}
\label{ampcp}
H_{22}(t_2),\ H_{22}(t_4),\ H_{22}(t_5),\ H_{22}(t_6),\ H_{22}(t_\text{peak})\,  	
\end{equation}
\item Frequency collocation points:
\begin{equation}
\label{freqcp}
\begin{split}
&\omega_{22}(t_1),\ \omega_{22}(t_2),\ \omega_{22}(t_3),\ \omega_{22}(t_4),\ \\ &\omega_{22}(t_5),\ \omega_{22}(t_6),\ \omega_{22}(t_\text{peak})\, 
\end{split}	
\end{equation}
\item Ringdown coefficients:
\begin{equation}
\label{rdcp}
d_{3},\  c_{3},\  c_{4},
\end{equation}
\item The time $t_{\text{MECO}}$ that corresponds to the MECO frequency.
\end{itemize}
where 
\begin{equation}
\begin{split}
&\{t_1,t_2,t_3,t_4,t_5,t_6\}=\\ &\{-10^5M,-2000M,-1000M,2t_{\text{MECO}} ,t_{\text{MECO}},0.25t_{\text{MECO}}\}.
\end{split}	
\end{equation}

\subsection{\label{sec:dataset}Dataset}

\begin{figure}
  \centering
  \includegraphics[width=.95\linewidth]{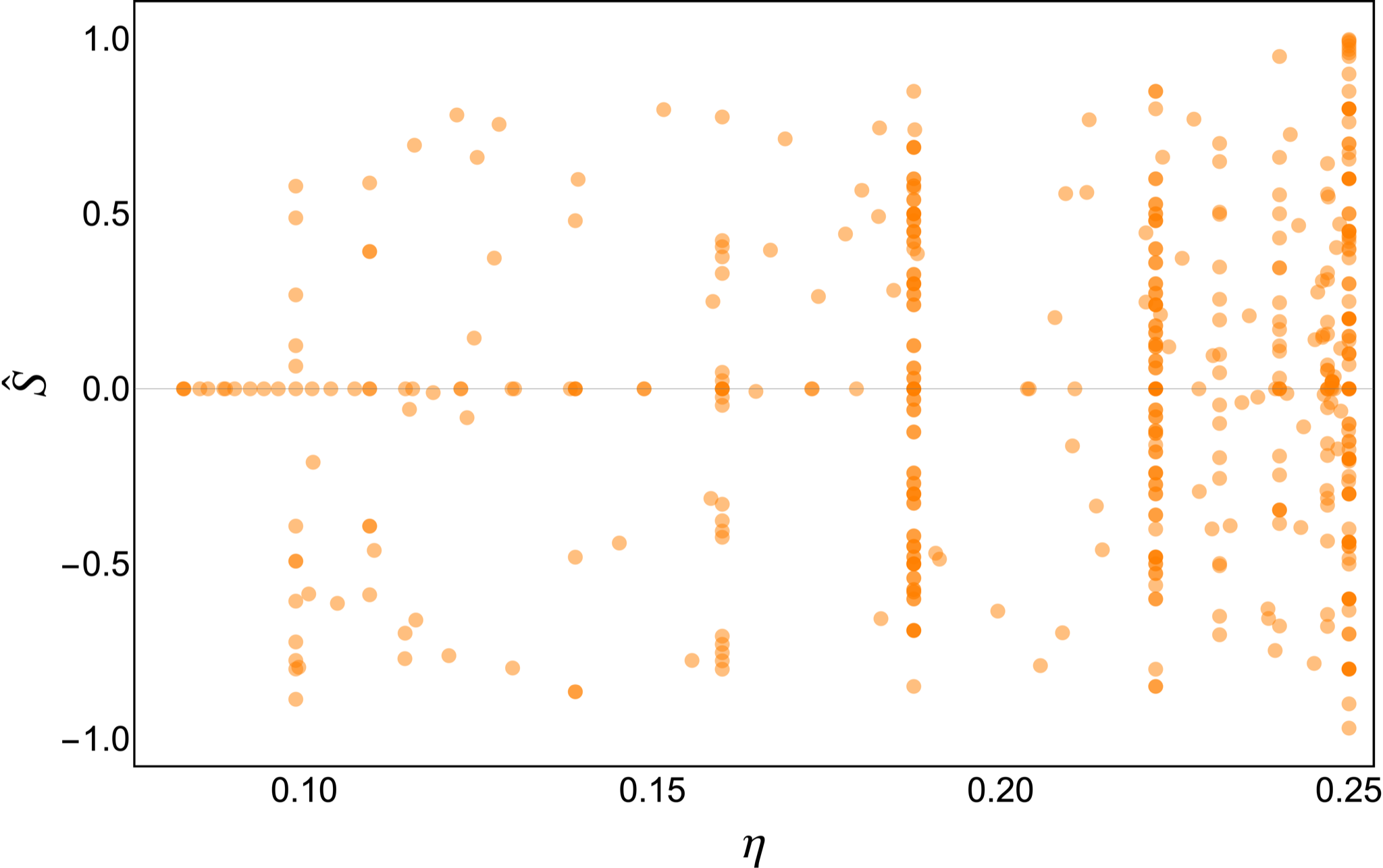}  
\caption{Parameter coverage of the non-precessing BBH simulations of the 2019 release of the SXS catalogue of Numerical Relativity Simulations.}
\label{fig:figParamspace}
\end{figure}

We have employed the latest release of the SXS Collaboration catalogue of numerical relativity simulations \cite{Boyle_2019} performed with the Spectral Einstein Code (SpEC), in particular the non-precessing quasi-circular set of simulations which comprises a total of 531 simulations. We have selected the highest available resolution level of the center-of-mass corrected extrapolated N=3 Regge-Wheeler-Zerilli strain for each simulation for the $l=2$, $m=2$ mode. The simulations span the 3D parameter space from $1\leq q\leq 10$ and $0\leq |\chi_{1,2}|\leq 0.998$ (see Fig.\,\ref{fig:figParamspace} for the parameter space coverage of the dataset). For the $\omega_{22}(t_1)$ collocation point, which is placed at $t=-10^5M$, effective-one-body (EOB) inspiral waveforms were computed using the $\text{SEOBNRv4}$ model at the same points in the parameter space as the numerical relativity simulations, for consistency. 

From the calibration dataset, 13 simulations were excluded because they appeared as outliers for some of the fitted quantities. In most of the cases, they are non-spinning simulations with parameters already covered by other simulations present in the catalogue. Other excluded simulations presented pathological behaviour, as SXS:BBH:0148. Their IDs and parameters are shown in Table \ref{tab:outliers}.

\begin{table}[h!]
\begin{center}
\begin{tabular}{|c|c|c|c|}
\hline
 SXS ID & q & $\chi_1$ & $\chi_2$ \\
 \hline
 \text{SXS:BBH:0002} & 1.00 & 0 & 0 \\
 \text{SXS:BBH:0040} & 3.00 & -0.50 & 0 \\
 \text{SXS:BBH:0148} & 1.0 & -0.44 & -0.44 \\
 \text{SXS:BBH:0149} & 1.00 & -0.20 & -0.20 \\
 \text{SXS:BBH:1110} & 7.00 & 0 & 0 \\
 \text{SXS:BBH:1111} & 5.00 & -0.90 & 0 \\
 \text{SXS:BBH:1142} & 1.25 & 0 & 0 \\
 \text{SXS:BBH:1145} & 1.25 & 0 & 0 \\
 \text{SXS:BBH:1362} & 1.00 & 0 & 0 \\
 \text{SXS:BBH:1363} & 1.00 & 0 & 0 \\
 \text{SXS:BBH:1369} & 2.00 & 0 & 0 \\
 \text{SXS:BBH:1370} & 2.00 & 0 & 0 \\
 \text{SXS:BBH:1374} & 3.00 & 0 & 0 \\
 \hline
\end{tabular}
\caption{\label{tab:outliers}List of SXS simulations excluded from the calibration dataset.}
\end{center}
\end{table}

\subsection{\label{sec:calibstrategy}Calibration strategy}

In order to fit these quantities across a three-dimensional parameter space and avoid both overfitting and underfitting, we employ the hierarchical fitting procedure presented in \cite{jimenez2017hierarchical}, which also has recently been employed to calibrate the IMRPhenomXAS \cite{pratten2020setting} and IMRPhenomXHM \cite{garcaquirs2020imrphenomxhm} frequency domain models. The method is based on constructing an appropriate ansatz through a sequence of fits to lower dimensional subspaces,  which are typically more densely populated with numerical simulations. First,  1-dimensional fits are performed for the dependence on the symmetric mass-ratio for the non-spinning subset of the dataset, and the spin dependence
of equal black holes. In order to capture in a simpler way the full two-dimensional spin-dependence, it is re-parameterized in terms of a dominant effective spin $\hat{S}$ and the spin difference $\Delta\chi$:

\begin{subequations}
    \begin{align}
    &\hat{S} = \frac{m_1^2\chi_1^2 + m_2^2\chi_2^2}{m_1^2+m_2^2},\\
    &\Delta\chi = \chi_1 - \chi_2,
    \end{align}
\end{subequations}
which were employed in the construction of the final state fits of \cite{jimenez2017hierarchical} and in some of the fits constructed for IMRPhenomXAS and IMRPhenomXHM models. For the dominant effective spin effects, one performs another 1-dimensional fit for a particular mass-ratio, which we set to $q=1$ for all the fits. With both 1D fits, one constrains a 2D fit over the equal spin subset. From this fit, residuals with the unequal spin cases are computed, and this residuals are fitted as a function of spin difference and mass-ratio in a domain $q\in[1,10]$. The resulting phenomenological fits of the quantities specified in (\ref{ampcp}), (\ref{freqcp}), (\ref{rdcp}) and the fit for the MECO time are included in a supplementary Mathematica package.

\section{\label{sec:validation}Validation}

In this section we assess the validity region of the model, the accuracy reproducing the dataset employed for the calibration of the model and its comparison to other aligned-spin models for the 22 mode, employing a common quantity called mismatch between two waveforms:
\begin{equation}
1-\mathcal{M}=1 - \max_{t_0,\phi_0}\dfrac{(h_1|h_2)}{\sqrt{(h_1|h_1)(h_2|h_2)}}.
\end{equation}
The inner product in the function space of waveforms is defined as:
\begin{equation}
(h_1|h_2)\equiv 4\Re\int_{f_{min}}^{f_{max}}df \dfrac{\tilde{h}_1(f)\tilde{h}_2(f)}{S_f(f)}
\end{equation}
where $\tilde{h}(f)$ is the Fourier transform of $h(t)$ and $S_f(f)$ is a frequency dependent weight with dimensions of time that typically is chosen as the estimated power spectral density of a particular configuration of the laser interferometric gravitational wave detectors. For the results presented below we use the zero noise high detuned power analytical PSD configuration \cite{adligopsd}.

\begin{figure}
  \centering
  \includegraphics[width=.95\linewidth]{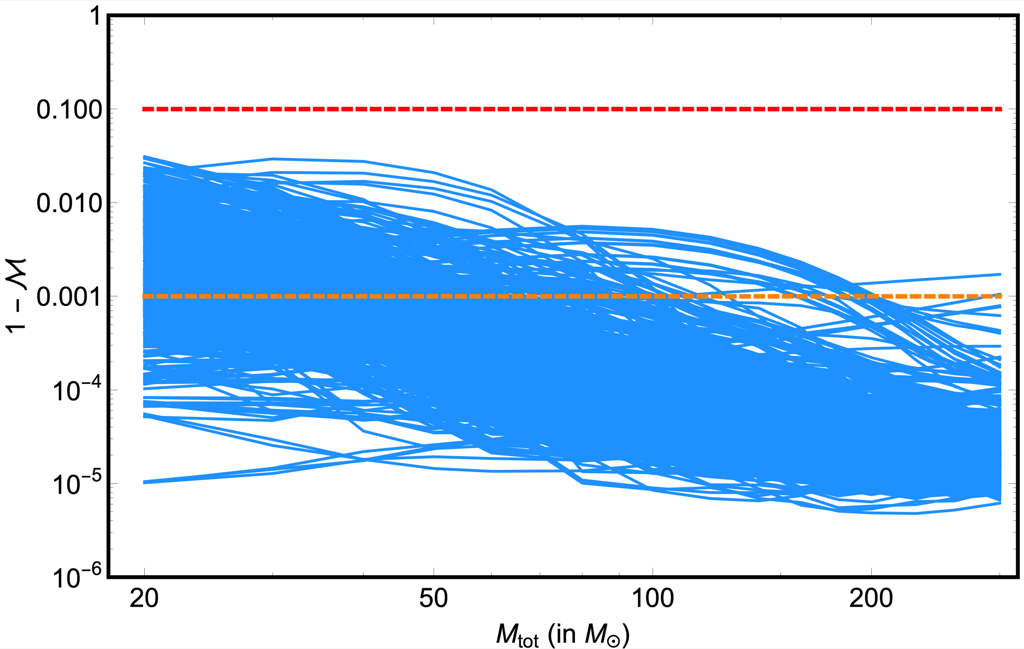}  
\caption{Mismatch between PhenomT and EOB-NR Hybrid catalogue constructed from the last SXS catalogue release.}
\label{fig:figMatchHyb}
\end{figure}

\begin{figure}
  \includegraphics[width=.75\linewidth]{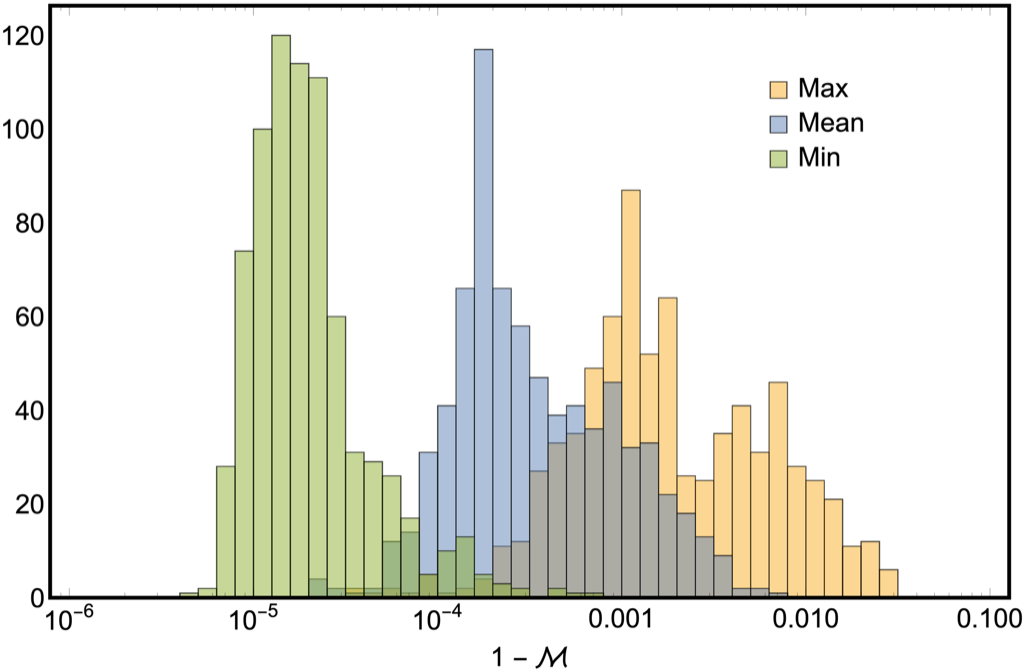}  
  \includegraphics[width=.75\linewidth]{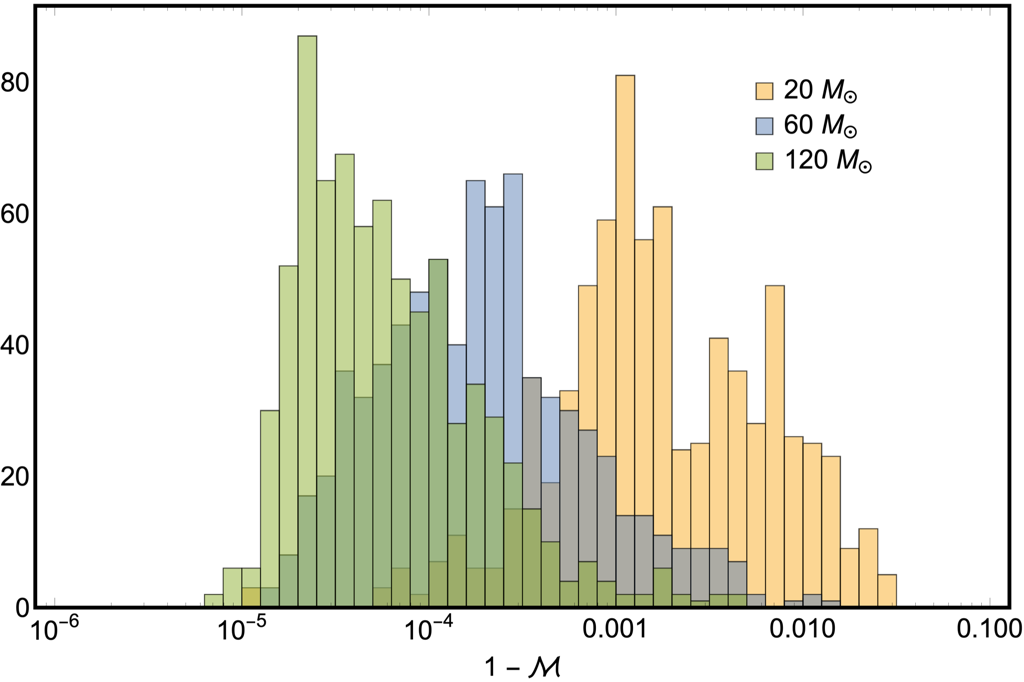}  
\caption{Histograms of mismatches between PhenomT and EOB-NR Hybrid catalogue. Top: distribution of maximum, mean and minimum mismatch for each case. Bottom: Distribution of mismatches at 20, 60 and 120 solar masses.}
\label{fig:figHistMatch}
\end{figure}

To assess the validity of the model with the dataset of numerical simulations employed in the calibration, we first compute the mismatch between the model and a set of EOB-NR hybrid waveforms contructed from the NR simulations of the dataset, in order to employ longer waveforms that are valid at least down to $20\text{Hz}$ at $20M_{\odot}$ (the same dataset of hybrids was employed in the calibration of IMRPhenomXAS and IMRPhenomXHM , see \cite{SaschaHybrids} and \cite{garcaquirs2020imrphenomxhm} for details on the construction of the hybrid waveform dataset). Mismatches were computed for different total mass $M=m_1 + m_2$ values between 20 and 300 $M_{\odot}$ for a minimum frequency of $20$ $\text{Hz}$ and a maximum frequency of $2048$ $\text{Hz}$, which correspond to the frequency band of typical BBH signals detectable by ground-based interferometers. In Fig.\,\ref{fig:figMatchHyb} we show the dependence of the mismatch on the total mass of the binary, the total mass scales the frequency such that for lower masses the waveform is shifted to higher frequencies and viceversa. We observe that the mismatches improve as we increase the total mass. In Fig.\,\ref{fig:figHistMatch} we show histograms of the mismatch
for particular values of the total mass, $(20, 60, 120) , M_{\odot}$, and the distribution of the minimal, maximal, and mean values of the computed mismatches. It can be seen that 
results improve as the mass increases.  We trace the degrading mismatches for lower masses to the inaccuracy of setting $t_0=0$ for TaylorT3 in the inspiral even with the inclusion of the early inspiral collocation point designed for mitigating this effect.

\begin{figure*}[htpb]
  \includegraphics[width=.3\linewidth]{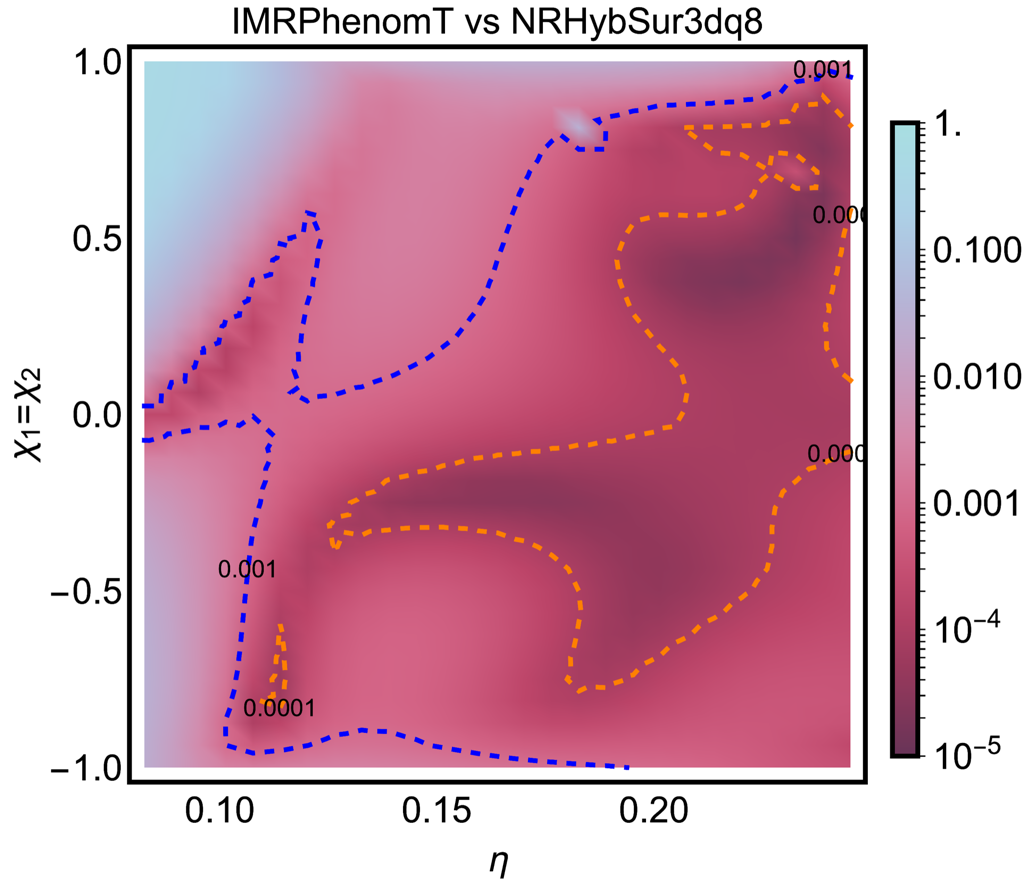}
  \includegraphics[width=.3\linewidth]{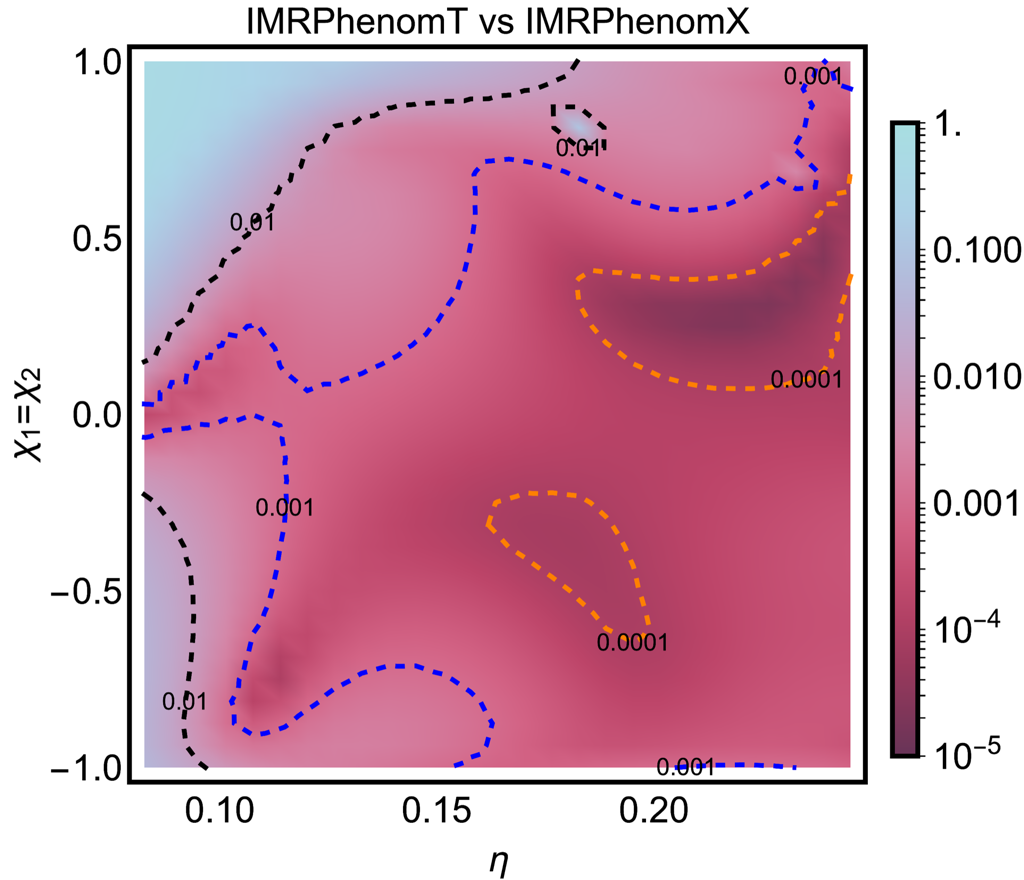}  
  \includegraphics[width=.3\linewidth]{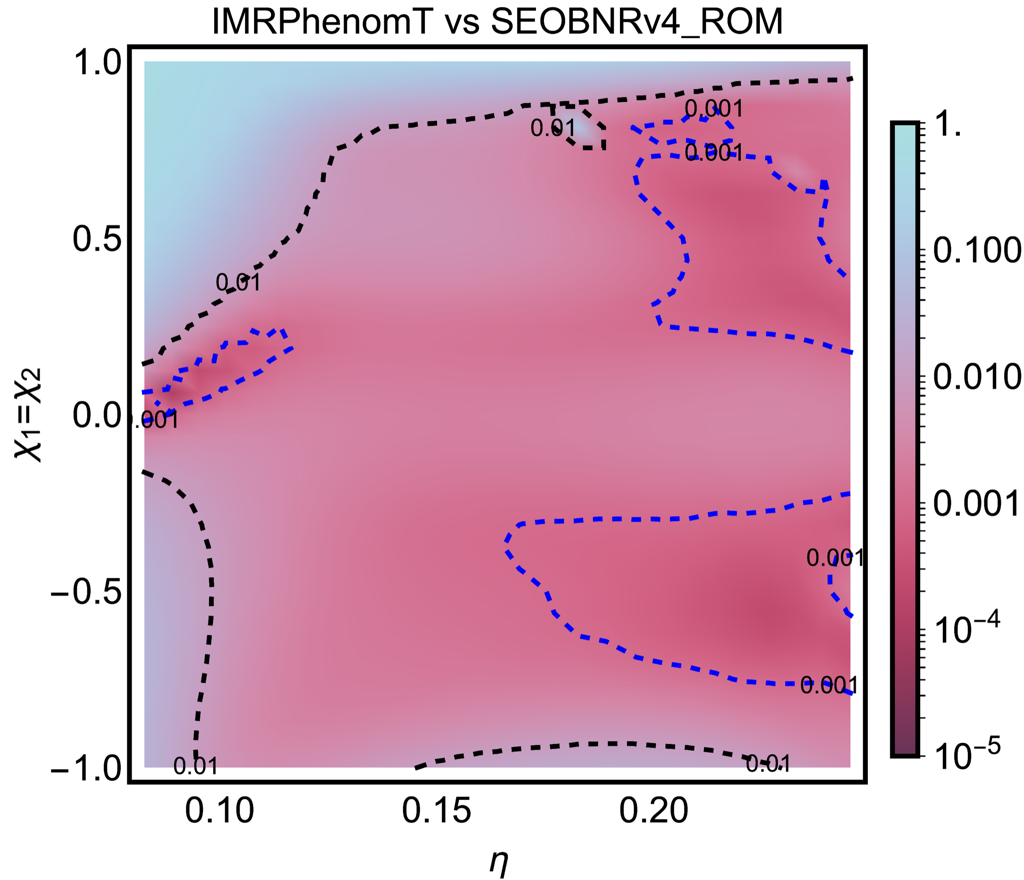}
 \includegraphics[width=.3\linewidth]{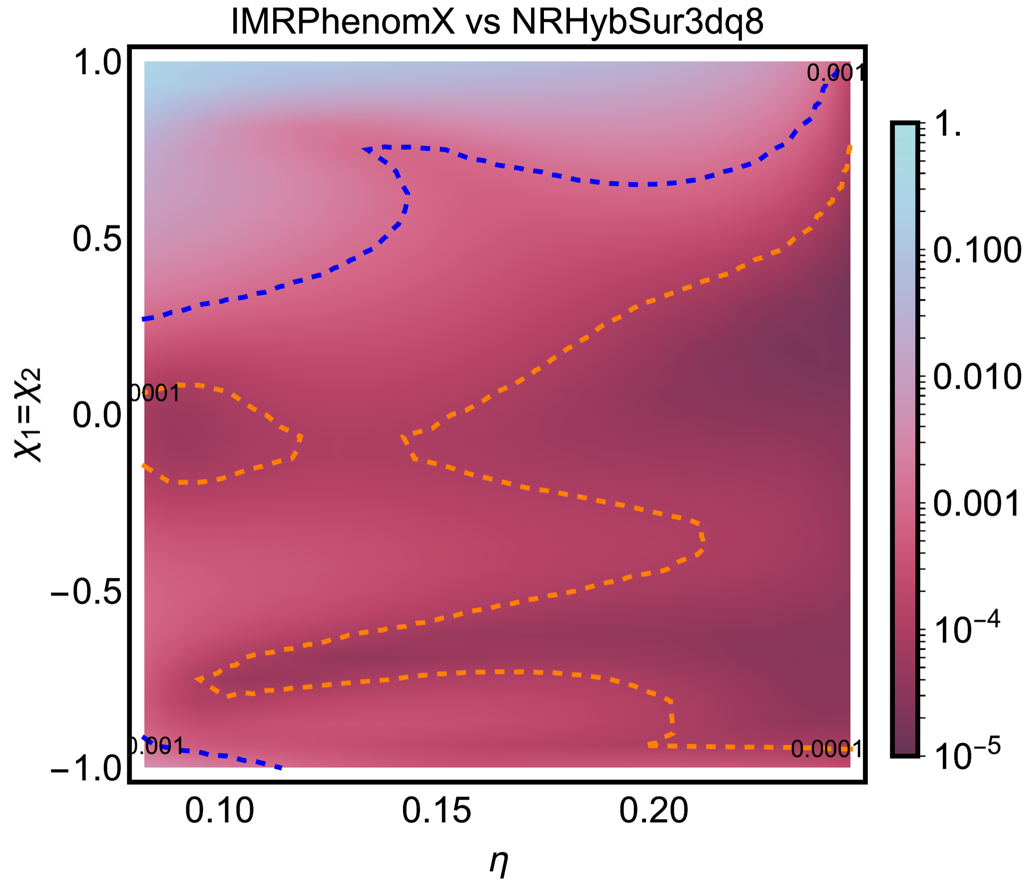}
 \includegraphics[width=.3\linewidth]{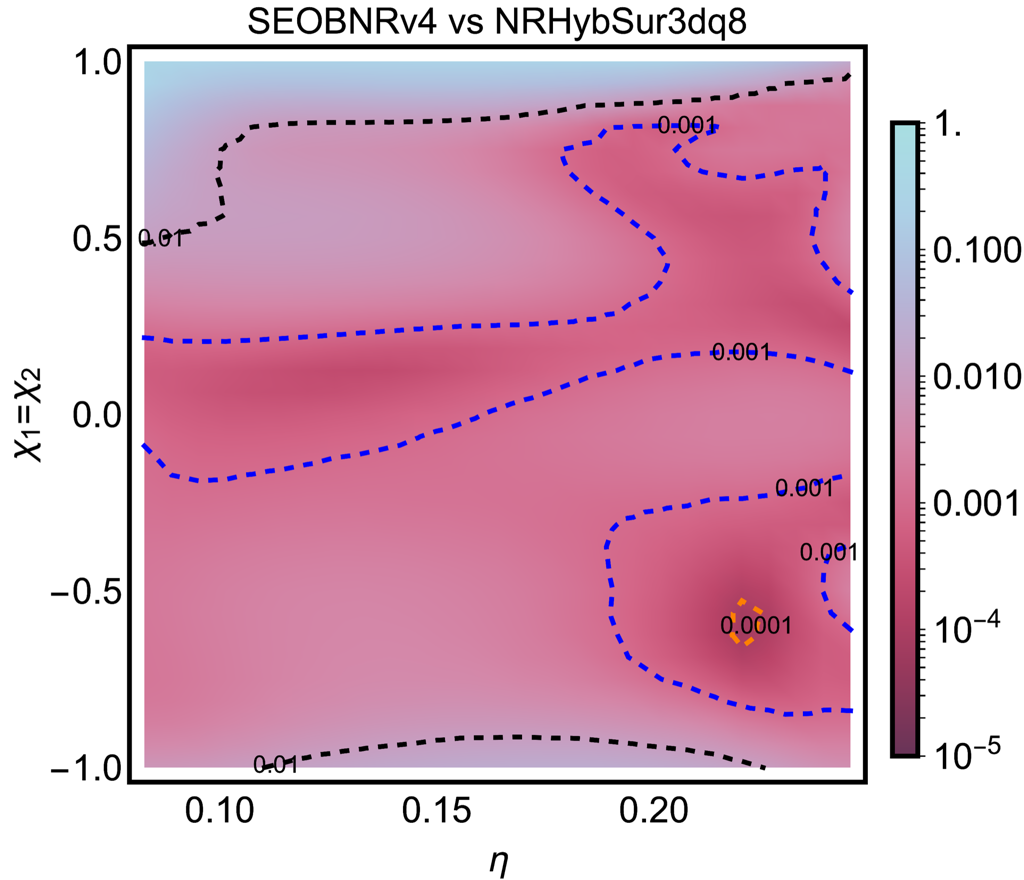}
 \includegraphics[width=.3\linewidth]{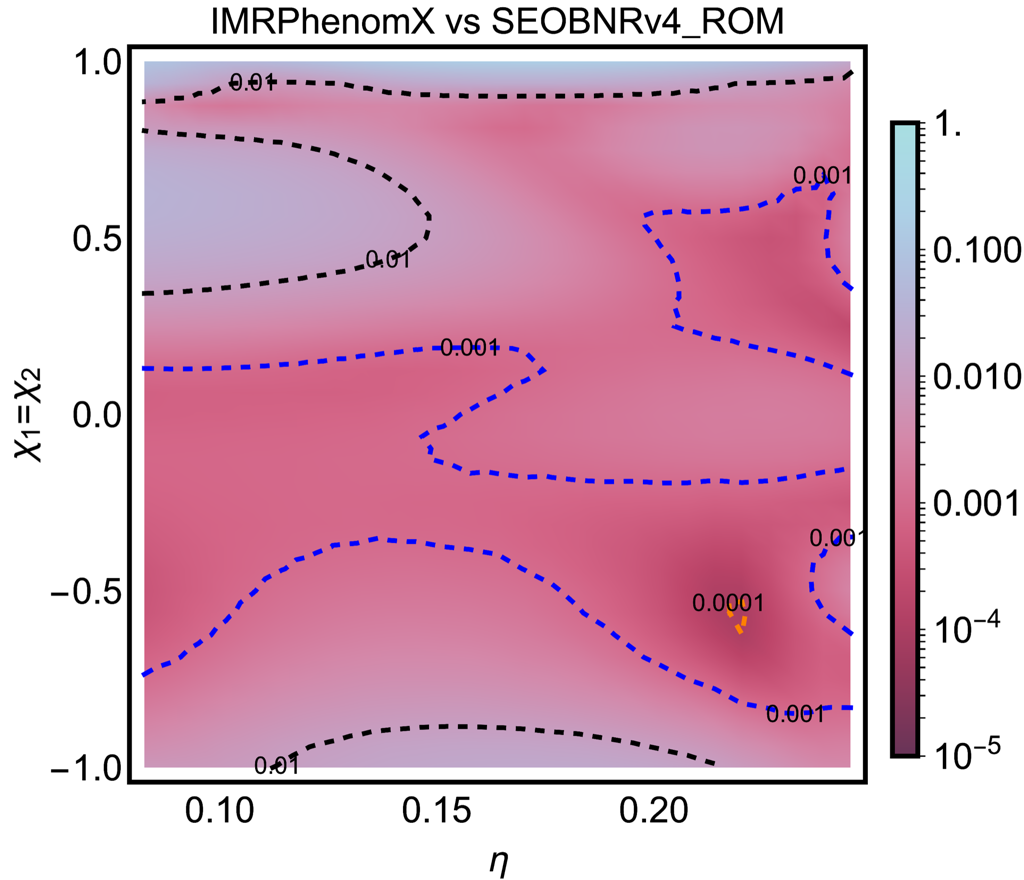}
 
\caption{Mismatch for $M=60 M_{\odot}$ between IMRPhenomT, IMRPhenomX, SEOBNRv4ROM and NRSurHyb3dq8 models. Black dashed lines: 1\% mismatch contour. Blue dashed lines: 0.1\% mismatch contour. Orange dashed lines: 0.01\% mismatch contour.}
\label{fig:figContour}
\end{figure*}

In addition to the comparison with numerical relativity waveforms we also compare our model to three state-of-the-art models for the non-precessing 22 mode, NRHybSur3dq8 \cite{Varma_2019}, IMRPhenomXAS \cite{pratten2020setting} and SEOBNRv4 \cite{bohe2017improved} in its ROM (reduced order model) version, for the equal spin 2D parameter space for a total mass of $60M_{\odot}$, since the hybrid mismatches have their mean value around this mass. In Fig.\,\ref{fig:figContour} we can observe that IMRPhenomT agrees better with IMRPhenomXAS than with SEOBNRv4, which is consistent with the fact that the calibration strategy and the dataset employed are similar for both IMRPhenom models. Also, both Phenom models agree with NRHybSur3dq8 better than SEOBNRv4. According to the results, the validity region of our model in which the discrepancy is below 1\% with the three models is for $q\leq 4$, if well for small spin magnitudes the model agrees up to $q=10$ and $q=8$ for negative spins.

\section{\label{sec:precessing}``Twisting up'' precessing extension}

We now turn to extending our model to precession. For describing a precessing quasi-circular binary, we need to extend the three-dimensional parameter space of aligned-spin configurations to the 
7-dimensional parameter space that includes all six individual spin components:
\begin{equation}
    \{\eta,\boldsymbol{\chi}_1,\boldsymbol{\chi}_2\},
\end{equation}
where the individual spin vectors $\boldsymbol{\chi}_{1,2}$ and the orbital angular momentum of the binary $\boldsymbol{L}$ can evolve due to the spin-orbit and spin-spin interactions, causing a precessional motion of the orbital plane.

A common approach for extending a non-precessing waveform model to describe precessing systems is to employ the ``twisting up'' procedure \cite{Schmidt_2012} based on the quadrupole alignment approximation \cite{Schmidt_2011}, \cite{PhysRevD.85.084003}, \cite{Boyle_2011}. The basic idea is that much of the precessional behavior can be captured by a time-dependent Euler rotation of the orbital plane. For a recent discussion of shortcomings of this approach see \cite{ramosbuades2020validity}.

\subsection{\label{sec:twistingup}``Twisting up" procedure''}


We follow the procedures of \cite{Boyle_2011, Schmidt_2011, Schmidt_2012, Schmidt_2015, PhysRevD.85.084003}. Defining a non-inertial reference frame that coprecesses with the orbital plane, the gravitational wave modes resemble those of a corresponding non-precessing system, i.e a system with the same mass ratio and spin magnitudes equal to the projection of the spin vectors onto the orbital angular momentum direction of the precessing system: 
\begin{equation}
    h^{\text{coprec}}_{lm}(t;\eta,\boldsymbol{\chi}_1,\boldsymbol{\chi}_2)\approx h^{\text{AS}}_{lm}(t;\eta,\chi_{1l},\chi_{2l}).
\end{equation}

The transformation from the coprecessing frame to an inertial reference frame corresponds to a time dependent instantaneous Euler rotation. The SWSH modes are spin-2 fields that transform under rotations as: 
\begin{equation}
    h_{lm}(t)=\mathcal{D}^{l}_{mm'}(\alpha,\beta,\gamma)h_{lm'}(t)\,
\end{equation}
where the $\mathcal{D}^{l}_{mm'}$ are the Wigner-D matrix elements:
\begin{equation}
    \mathcal{D}^{l}_{mm'}(\alpha,\beta,\gamma)=e^{im\alpha}e^{im'\gamma}d^{l}_{mm'}(\beta),
\end{equation}
and $d^{l}_{mm'}$ are the Wigner-d matrices defined as
\begin{subequations}
    \begin{equation}
    \begin{split}
       & d^{l}_{mm'}(\beta) = \\
       & \sum_{k=k_{\text{min}}}^{k_{\text{max}}}c_{l,m,m',k}\Big(\cos{\dfrac{\beta}{2}}\Big)^{2l+m-m'-2k}\Big(\sin{\dfrac{\beta}{2}}\Big)^{2k-m+m'},
    \end{split}
    \end{equation}
    \begin{equation}
    \begin{split}
       & c_{l,m,m',k} = \\
       & \dfrac{(-1)^k}{k!}\dfrac{\sqrt{(l+m)!(l-m)!(l+m')!(l-m')!}}{(l+m-k)!(l-m'-k)!(k-m+m')!}.
    \end{split}
    \end{equation}
\end{subequations}
Most of the precessing dynamics of the system are encoded in the rotation Euler angles $\alpha$, $\beta$ and $\gamma$.

The quadrupole-aligned frame can be defined in terms of the direction of maximum emission of the binary system, that can be computed analytically from the modes in the inertial frame. Specifying this direction as $\hat{\boldsymbol{\ell}}$, the Euler rotation angles are defined as:
\begin{subequations}
\label{eq:eulerangles}
\begin{equation}
    \alpha=\arctan(\ell_y,\ell_x),
\end{equation}
\begin{equation}
    \cos\beta=\hat{\boldsymbol{J}}\cdot\hat{\boldsymbol{\ell}}=\ell_z,
\end{equation}
\end{subequations}
while the freedom in the third angle can be fixed imposing the minimal rotation condition \cite{Boyle_2011}:
\begin{equation}
\label{eq:minimalcond}
\dot{\gamma}=-\dot{\alpha}\cos\beta\ .
\end{equation}
Finding the time dependence behaviour of the Euler angles requires then to know the evolution of the maximum emission direction vector $\hat{\boldsymbol{\ell}}$. 


\subsection{\label{sec:sec5subsec2}Precessing angles}

When the precessional timescale is much greater than the orbital timescale, precessing variation of the different momenta can be averaged per orbit, giving the following set of equations up to 2PN relative order (\cite{PhysRevD.31.1815}, \cite{Barker}, \cite{Buonanno_2006}):
\begin{subequations}
\label{eq:preceq1}
\begin{align}
\dot{\hat{\boldsymbol{L}}}  =& \Big\{\Big(2 + \dfrac{3}{2}q\Big)-\dfrac{3}{2}\dfrac{v}{\eta}\Big[(\boldsymbol{S}_2 + q\boldsymbol{S}_1)\cdot\hat{\boldsymbol{L}}\Big]\Big\}v^6(\boldsymbol{S}_1\times\hat{\boldsymbol{L}})\nonumber\\
&+\Big\{\Big(2 + \dfrac{3}{2q}\Big)-\dfrac{3}{2}\dfrac{v}{\eta}\Big[(\boldsymbol{S}_1 + \dfrac{1}{q}\boldsymbol{S}_2)\cdot\hat{\boldsymbol{L}}\Big]\Big\}v^6(\boldsymbol{S}_2\times\hat{\boldsymbol{L}}  )\nonumber\\
&+\mathcal{O}(v^7),\\
\dot{\boldsymbol{S}}_1  =& \Big\{\eta\Big(2 + \dfrac{3}{2}q\Big)-\dfrac{3v}{2}\Big[(\boldsymbol{S}_2 + q\boldsymbol{S}_1)\cdot\hat{\boldsymbol{L}}\Big]\Big\}v^5(\hat{\boldsymbol{L}} \times\boldsymbol{S}_1 )\nonumber\\
&+  \dfrac{v^6}{2}\boldsymbol{S}_2\times\boldsymbol{S}_1+\mathcal{O}(v^7),\\
\dot{\boldsymbol{S}}_2  =& \Big\{\eta\Big(2 + \dfrac{3}{2q}\Big)-\dfrac{3v}{2}\Big[(\boldsymbol{S}_1 + \dfrac{1}{q}\boldsymbol{S}_2)\cdot\hat{\boldsymbol{L}}\Big]\Big\}v^5(\hat{\boldsymbol{L}} \times\boldsymbol{S}_2 )\nonumber\\
&+  \dfrac{v^6}{2}\boldsymbol{S}_1\times\boldsymbol{S}_2+\mathcal{O}(v^7),
\end{align}
\end{subequations}

where the individual spin magnitudes $S_1$ and $S_2$ and the spin projections onto the orbital angular momentum direction $S_{1z}$ and $S_{2z}$ are conserved at 2PN order, the spin directions $\hat{\boldsymbol{S}}_1$ and  $\hat{\boldsymbol{S}}_1$ and correspondingly $\hat{\boldsymbol{L}}$ vary in the precessing timescale, $L_z$ and $J$ vary in the radiation-reaction timescale and the total angular momentum direction $\hat{\boldsymbol{J}}$ is approximately a conserved quantity \cite{Racine_2008}.

In terms of a time-dependent Euler rotation of $\boldsymbol{L}$, $\boldsymbol{S}_1$ and $\boldsymbol{S}_2$ around the quasi-conserved direction $\hat{\boldsymbol{J}}$, evolution equations for the Euler angles $\alpha$, $\beta$ and $\gamma$ are obtained. 

\subsubsection{Next-to-next-to-leading order precessing average single spin approach}

Introducing the triad $\{\boldsymbol{n},\boldsymbol{\lambda},\boldsymbol{\ell}\}$, where $\boldsymbol{n}$ is the unit separation vector between both black holes, $\boldsymbol{l}$ is the direction of the unit vector normal to the instantaneous plane and $\boldsymbol{\lambda}$ completes the triad following the right hand rule $\boldsymbol{\lambda}=\boldsymbol{\ell}\times \boldsymbol{n}$, the evolution equations for the Euler angles in the single spin case are:
\begin{subequations}
\begin{align}
    \frac{d \alpha}{d t} &= -\frac{\bar{\omega}}{\sin \beta} \frac{J_n}{\sqrt{J^2_n + J^2_{\lambda}}} , \\
    \frac{d \beta}{d t} &= \bar{\omega} \frac{J_{\lambda}}{\sqrt{J^2_n + J^2_{\lambda}}}, \\
    \frac{d \gamma}{d t} &= - \dot{\alpha} \cos \beta ,
\end{align}
\end{subequations}
where $J_{n,\lambda,\ell}$ are the components of the total angular momentum $\boldsymbol{J}=\boldsymbol{L} + \boldsymbol{S}_1$ in this triad. These equations were solved to next-to-next-to-leading order in the spin-orbit coupling \cite{Boh__2013} and were employed in the IMRPhenomP and IMRPhenomPv2 models \cite{Schmidt_2015, Bohe:PPv2} and are also included in the IMRPhenomXP model \cite{phenomxphm}.

\subsubsection{Multiscale analysis double spin approach}

Motivated by the separation in timescales between radiation-reaction and precessing effects, in \cite{Chatziioannou_2017} the authors construct a perturbative solution to the orbit averaged spin evolution equations (\ref{eq:preceq1}) based on the multiscale analysis (MSA) technique and the known analytical solution for the conservative dynamics approximation \cite{Kesden_2015}. Here we very briefly summarize their method and results.
For the precessing angle $\alpha$, the leading MSA term is obtained by averaging over a precessing orbit, given that the precessing timescale is much faster than the radiation reaction timescale, giving a secular term $\alpha_{-1}$ that varies on the radiation-reaction timescale, and then a first correction $\alpha_0$ is computed varying in the precessing timescale, which introduces modulations to the secular term:

\begin{equation}
\alpha(t) = \alpha_{-1}(t)+\alpha_0(t).
\end{equation}
In order to compute the first correction, the time variation of the total spin vector $\boldsymbol{S}=\boldsymbol{S}_1 + \boldsymbol{S}_2$ has to be considered. Starting from the evolution equations (\ref{eq:preceq1}), the evolution equation for $S^2$
can be expressed as a third order polynomial in $S^2$ that varies only in the radiation-reaction time scale, and a expression for $S^2$ can be obtained in terms of the roots $S^2_{+}, S^2_{-}, S^2_{3}$ of the polynomial:
\begin{equation}
\label{eq:stoteq}
S^2=S^2_{+} + (S^2_{-}-S^2_{+})\text{sn}^2(\psi,m),
\end{equation}
where $\text{sn}(\psi,m)$ is a Jacobi Elliptic function with phase:
\begin{equation}
\label{eq:psidot}
\dot{\psi}(t)=\frac{A}{2}\sqrt{S^2_{+}-S^3_{+}}
\end{equation}
and parameter 
\begin{equation}
m=\dfrac{S^2_{+}-S^2_{-}}{S^2_{+}-S^2_{3}}.
\end{equation}
In the conservative dynamics approximation, $\dot{\psi}$ is constant and $\psi$ can be obtained directly. In the presence of radiation-reaction, the equation (\ref{eq:psidot}) can be integrated to 1 PN giving:
\begin{equation}
\label{eq:psiangle}
\psi=\psi_0-\dfrac{3g_0}{4}\delta m \, v^{-3}(1 + \psi_1 v + \psi_2 v^2),
\end{equation}
where $\psi_0$ is an integration constant that can be computed inverting equation (\ref{eq:stoteq}) at the reference time, $\psi_1$ and $\psi_2$ are constants that depend on the conserved quantities and $g_0=5/(96\eta)$ is the first PN coefficient of the $v$ evolution.
The opening angle $\beta(t)$ is defined as
\begin{equation}
\label{eq:betaMSA}
\cos\beta(t)=\hat{\boldsymbol{J}}\cdot\hat{\boldsymbol{L}}=\dfrac{J^2(t)+L^2(t)-S^2(t)}{2J(t)L(t)},
\end{equation}
and can be computed using equation (\ref{eq:stoteq}) for the total spin magnitude, a PN description for the evolution of the orbital angular momentum magnitude $L$ and computing the precessing averaged $J$ as:
\begin{equation}
J^2=L^2 + \dfrac{2c_1}{v} + \big\langle S^2\rangle _{\text{pr}}+\mathcal{O}(v)
\end{equation}
where $c_1$ depends on the initial values:
\begin{equation}
c_1 = \frac{v_0}{2}(J_0^2 - L_0^2 - \big\langle S^2\rangle _{\text{pr},0})
\end{equation}
Finally, the third Euler angle can be computed from eq. (\ref{eq:minimalcond}).

\subsubsection{Numerical integration of the spin evolution equations}

Besides the analytical NNLO and MSA approximations for the Euler angles, which we have summarized in the previous subsections, the model presented in this work also incorporates two numerical ways of obtaining the Euler angles. The first one is to numerically integrate the 2PN orbit averaged set of equations (\ref{eq:preceq1}), from which the MSA analytical approximation is derived, using the non-precessing gravitational wave frequency for computing the PN velocity $v(t)=(\omega_{22}(t)/2)^{1/3}$. This leads to a numerical solution of $\hat{\boldsymbol{L}}(t)$ from which the Euler angles can be computed from eqs. (\ref{eq:eulerangles}). The second one is to employ the public code PNEvolveOrbit, available in the LIGO Algorithm Library (LAL) software package to evolve the spin evolution equations using the SpinTaylorT4 approximant for the orbital evolution and with the capability of selecting the PN order of the spin-orbit and spin-spin interacting terms \cite{SturaniT4}.

\begin{figure}
 \centering
  \includegraphics[width=.95\linewidth]{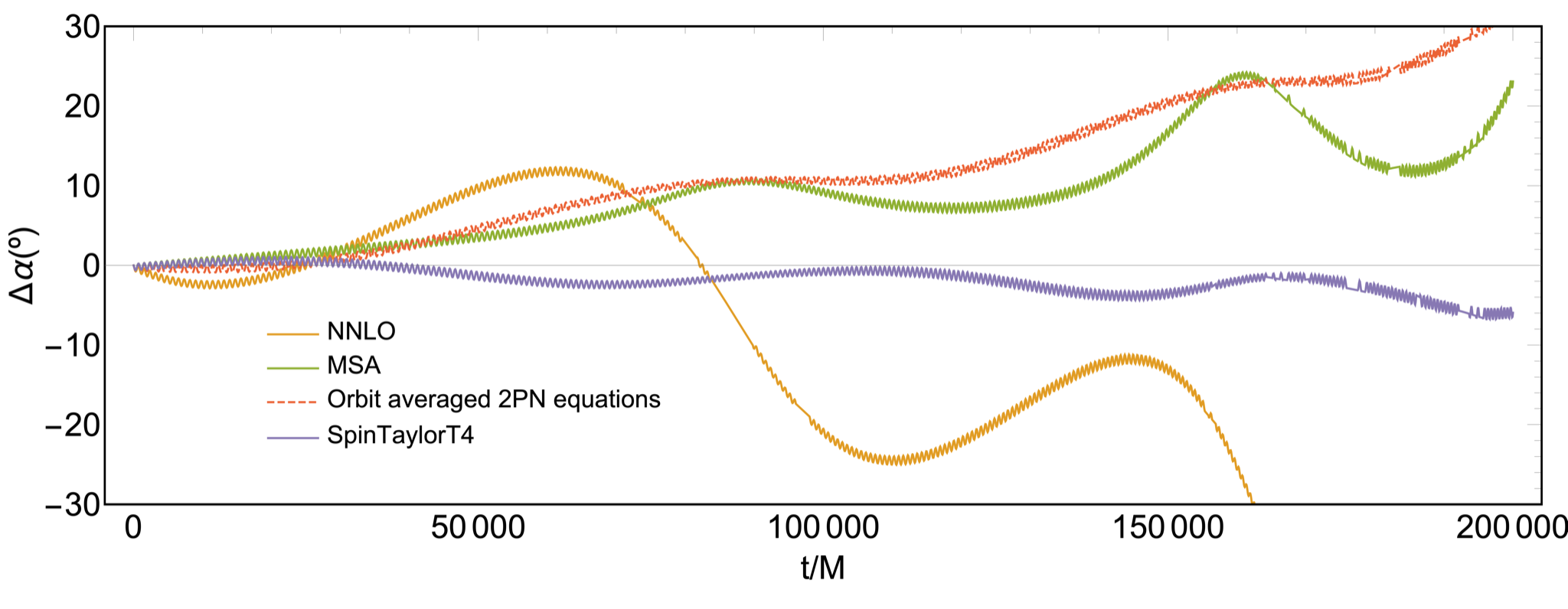}  
  \includegraphics[width=.95\linewidth]{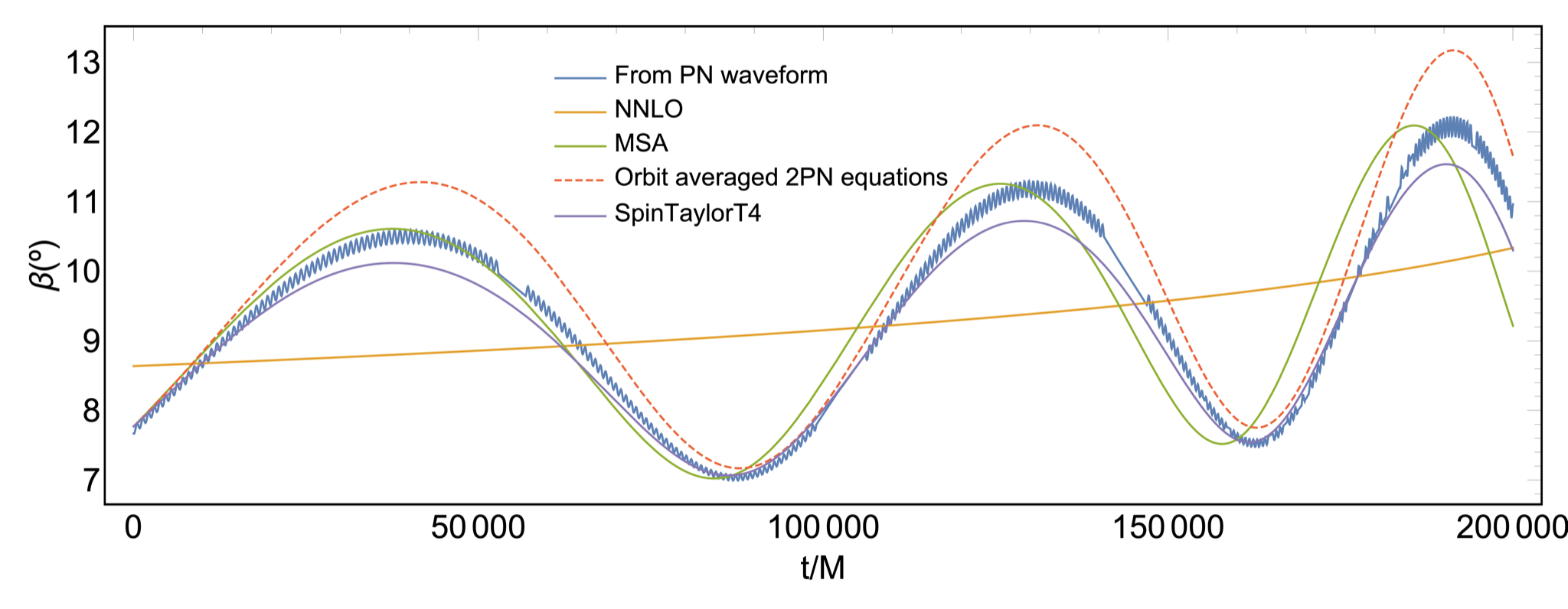}  
\caption{Comparison of the different options for the Euler angles $\alpha$ and $\beta$ with the Euler angles obtained with a full numerical PN evolution for parameters $q=2$, $\boldsymbol{\chi}_1=(-0.3,0.4,-0.2)$, $\boldsymbol{\chi}_2=(0.5,0,0.3)$ }
\label{fig:anglesPN}
\end{figure}

The main differences between the two numerical approaches are the presence of higher order PN corrections in the LAL function and the addition of in-plane spin contribution to the orbital angular momentum $\boldsymbol{L}$ which direction will not coincide completely with the Newtonian direction $\hat{\boldsymbol{\ell}}$. However, the LAL function employs for the orbital phase evolution SpinTaylorT4, which close to merger will become inaccurate as the conditions for the PN approximation start to fail. In a future implementation of the model in the LAL libraries, the spin evolution code of PNEvolveOrbit will be augmented with the option of employing PhenomT phase which is accurate at merger.

A systematic comparison of the different approaches to compute the Euler angles is out of the scope of this work and will be the subject of future work. Nevertheless, the treatment of a set of examples in the PN regime gives some general considerations and in Fig.\,\ref{fig:anglesPN} is shown a particular example. Both the NNLO and the MSA approximations seem to reproduce the secular growth of the angles at low frequencies, however this secular growth diverges with respect to a full numerical PN treatment as the system is evolving, being the MSA angles more accurate in reproducing the secular growth. With respect to the oscillations produced by the variation of the total spin in-plane for double spin systems, the NNLO approach loses this information by construction, while the MSA approach can track the correct phasing of the oscillations at low frequencies, starting to dephase at frequencies lower than the beginning of the NR regime. Both numerical approaches correctly reproduce  the phases in the oscillations, being the major difference between them that the LAL function PNEvolveOrbit is able to reproduce better the secular growth of the angles.

\subsubsection{Norm of the orbital angular momentum}

In the PN expressions describing the opening angle $\beta$, a description for the norm of the orbital angular momentum $L$ is needed. We incorporate a PN description up to 4PN order with spin-orbit interactions \cite{Le_Tiec_2012, Boh__2013, Damour_2014, Blanchet_2017, Bernard_2018}  allowing to select the desired expansion order, in a similar way as it is implemented in IMRPhenomXPHM \cite{phenomxphm}:
\begin{equation}
    \label{eq:Lorb}
    L(v(t))=\frac{\eta}{v}\sum_{n=0}^{8}l_n v^n(t),
\end{equation}
where the coefficients $l_{n}$ are shown in Appendix \ref{sec:appendixL}. For example, selecting up to 2PN with no spin contributions allows a more direct comparison with IMRPhenomPv2 \cite{Hannam_2014} while IMRPhenomPv3 \cite{Khan_2019} employs 3PN including spin-orbit.

\begin{figure}
  \centering
  \includegraphics[width=.95\linewidth]{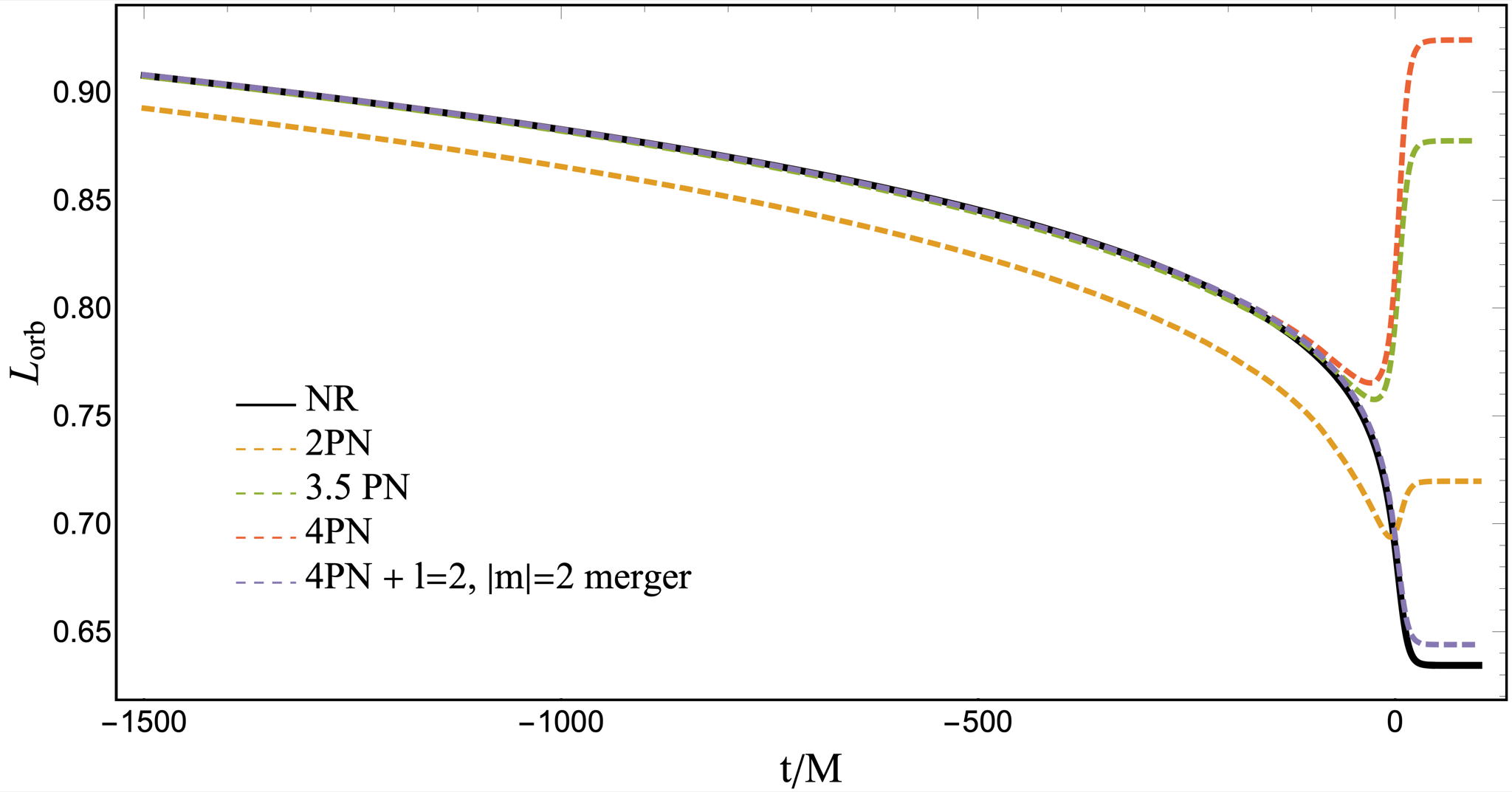}  
\caption{Comparison of orbital angular momentum computed from NR simulation SXS:BBH:0237 [$q=2$, $\chi_1=-0.6$, $\chi_2=0.6$] with PN at different expansion orders and the hybrid version implemented in the model.}
\label{fig:figLorb}
\end{figure}

While the PN descriptions at 3.5 and 4 PN order are accurate during the inspiral, they become inaccurate as the merger is approached, including the breaking of the monotonic behaviour for some cases. The approach that has been followed to improve the merger description in this work is to join the accurate 4 PN order inspiral description with the numerical computation of the wave radiated orbital angular momentum employing the underlying nonprecessing model for the $l=2$, $m=2$ mode:

\begin{equation}
\label{eq:Lcustom}
  L(t) =
  \begin{cases}
                                   L_{\text{4PN}}(t) & t\leq 2 \, t_{\text{MECO}} \\
                                   
                                   \\ L_{\text{4PN}} (2 \, t_{\text{MECO}}) -  L^{22}_{\text{rad}}(t) & 2 \, t_{\text{MECO}}\leq t,
  \end{cases}  
\end{equation}
where
\begin{subequations}
\begin{align}
    L^{22}_{\text{rad}}&\simeq \int dt \dot{J}^{22}_z,\\
    \dot{J}^{22}_z&=\lim_{r\rightarrow\infty}\frac{r^2}{8\pi}\text{Im}\left\{h_{22}(t)\dot{\bar{h}}_{22}(t)-h_{2,-2}(t)\dot{\bar{h}}_{2,-2}(t)\right\}
\end{align}
\end{subequations}
and the expression for $\dot{J}^{22}_z$ comes from the general expression in \cite{Ruiz_2007} and the approximation holds when the emission of the individual spin components is negligible. In Fig.\,\ref{fig:figLorb} a comparison of $L$ at different PN orders and the hybrid construction (\ref{eq:Lcustom}) with a NR simulation is presented, illustrating the problems of the PN descriptions in the merger region and the validity of the proposed approximate solution. 
Neglecting the subdominant harmonics contribution to the radiated angular momentum, which become more important as the mass-ratio increases, is an important caveat in our proposed solution as currently implemented. However, this will be mitigated with a future extension of the model by the treatment of subdominant harmonics.

\subsection{Final state}

One of the limitations of the twisting up procedure for mapping non-precessing waveforms into precessing ones is the final state emission, i.e the ringdown signal. In general, the final state of a precessing binary and the final state of its analog non-precessing counterpart will be different. Since the ringdown signal depends on the final mass and spin of the remnant black hole, the map will be inaccurate in this region. A common fix employed by phenomenological waveform models is to construct the coprecessing modes from the aligned-spin model employing the corresponding final state of the desired precessing waveform.

However, in the precessing situation,  phenomenological fits of the final state quantities in the 7-dimensional parameter space are not yet available. Numerical fits have been performed using Gaussian Process Regression techniques \cite{Varma_2019_Remnant}, and alternatives employing Deep Neural Network have also been presented recently \cite{haegel2019predicting}. For PhenomTP, we choose to employ the same approach as other phenomenological models as \cite{Hannam_2014, Khan_2019, phenomxphm} and approximate the final spin of the precessing system by an in plane spin augmentation of the non-precessing final spin fit:
\begin{equation}
    \chi_f^{\text{augmented}}=\sqrt{\chi_{f,AS}^2 + S_{\perp}^2/M^4}
\end{equation}
where $\boldsymbol{S}_{\perp}$ is the in-plane total spin at some frequency. Ideally, this frequency should correspond to the merger frequency, where the final black hole is formed. 

However, the PN descriptions of the Euler angles, from which one could compute the total spin in-plane, are not accurate enough to predict $S_{\perp}$ at this frequency. In particular, in the analytical approaches, the NNLO approximation does not even carry double spin information and we have seen that the MSA double spin effects dephase at late inspiral. Different ways of addressing the definition of $S_{\perp}$ in the final spin formula can be tested. For IMRPhenomPv2, which employs the NNLO description of the Euler angles, the $\chi_p$ quantity is employed, which in some sense averages between the maximum and minimum values of the in-plane total spin. From the construction of the MSA angles, the phase $\psi$ (eq.~\ref{eq:psiangle}) that regulates the norm of the total spin between the roots $S_{+}$ and $S_{-}$ can be employed, if well PN comparisons have shown that may not be accurate enough to predict the correct behaviour at merger, so in some sense the merger value will be randomized. The numerical evolution equations, however, could predict a better estimate for the merger total spin direction, at least the total spin direction is tracked better during a larger duration of the inspiral, if well the behaviour at merger should be tested. Currently, the model only implements a somehow arbitrary decision of computing the total in-plane spin at the reference frequency, but we emphasize that this is not the optimal choice and the testing and selection of a better approach will be studied in future work, which will benefit not only the model presented in this work, but also other models from the Phenom family. A related discussion is present in the final spin section of \cite{phenomxphm}, since different ways of computing the final spin  are also implemented in the IMRPhenomXPHM model.

Another problem in the ringdown description is due to the inaccuracy of the PN description for the Euler angles on this region. For example, the NNLO description, which essentially depends on the gravitational wave frequency and not in its derivative, gives no evolution after the merger, while it is known from NR simulations that the final black hole suffers an effective precessional motion \cite{O_Shaughnessy_2013}. A simple approximation that shows the time dependence of the Euler angles during the ringdown is to take the leading contribution for small opening angle of the twisting up formula considering only the twisting of the $l=2,|m|=2$ coprecessing modes:
\begin{equation}
    h^{P}_{2m}\simeq e^{-im\alpha}e^{i2\epsilon}d^2_{2m}(\beta)h^{\text{coprec}}_{22},
\end{equation}
and compute the complex ratio between the inertial $m=2$ and $m=1$ modes:
\begin{equation}
\label{eq:ratiord}
    h^{P}_{22}/h^{P}_{21}\simeq -\frac{1}{2}e^{-i\alpha}\tan(\beta/2).
\end{equation}
Expressing the modes in the ringdown as a superposition of QNM states and considering only the leading ground state:
\begin{equation}
    h_{2m}^{\text{RD}}\simeq H_0 e^{-\omega^{\text{damp}}_{12m}}e^{i\omega^{\text{RD}}_{12m}},
\end{equation}
and employing equation (\ref{eq:ratiord}), the leading contribution to the Euler angles $\alpha$ and $\beta$ (and then $\gamma$ employing the minimal rotation condition (\ref{eq:minimalcond})) during the ringdown is:
\begin{subequations}
\label{eq:anglesrd}
\begin{align}
    \alpha^{\text{RD}}(t)&\simeq (\omega^{\text{RD}}_{122}-\omega^{\text{RD}}_{121})t + \alpha_0^{\text{RD}},\\
    \beta^{\text{RD}}(t)&\simeq -2\arctan\Big(2e^{(\omega^{\text{damp}}_{121}-\omega^{\text{damp}}_{122})t}\Big)+\beta_0^{\text{RD}}.
\end{align}
\end{subequations}
The result for the $\alpha$ angle is consistent with a similar derivation of this approximation done in \cite{marsat2018fourierdomain} and it is also implemented in the precessing EOB models SEOBNRv4P and SEOBNRv4PHM \cite{ossokine2020multipolar}. Phenomenologically, it can be observed that the precessing angle $\alpha$ will increase during the ringdown if $\omega^{\text{RD}}_{122}>\omega^{\text{RD}}_{121}$ (which is typically the case) and the opening angle $\beta$ will tend to open or to close depending on the relation between $\omega^{\text{damp}}_{122}$ and $\omega^{\text{damp}}_{121}$ which is more case dependent.

The PN description of the angles cannot afford for this behaviour, since its dependence on the orbital frequency, that can be approximated by half the $l=2$, $m=2$ wave frequency, produces a stationary value when the ringdown frequency is achieved. At the current stage, the leading order ringdown contributions to the precessing Euler angles are implemented as an option in the model, but with a simple implementation consisting in cutting the inspiral angle descriptions at the peak time and then joining the ringdown approximation.  It will be studied in more detail in future work better approaches to connect the inspiral and ringdown descriptions for the angles, with the aim of not only improving the model presented in this paper, but also to provide a solution for the frequency domain phenomenological models.

\begin{figure}
 \centering
  \includegraphics[width=.85\linewidth]{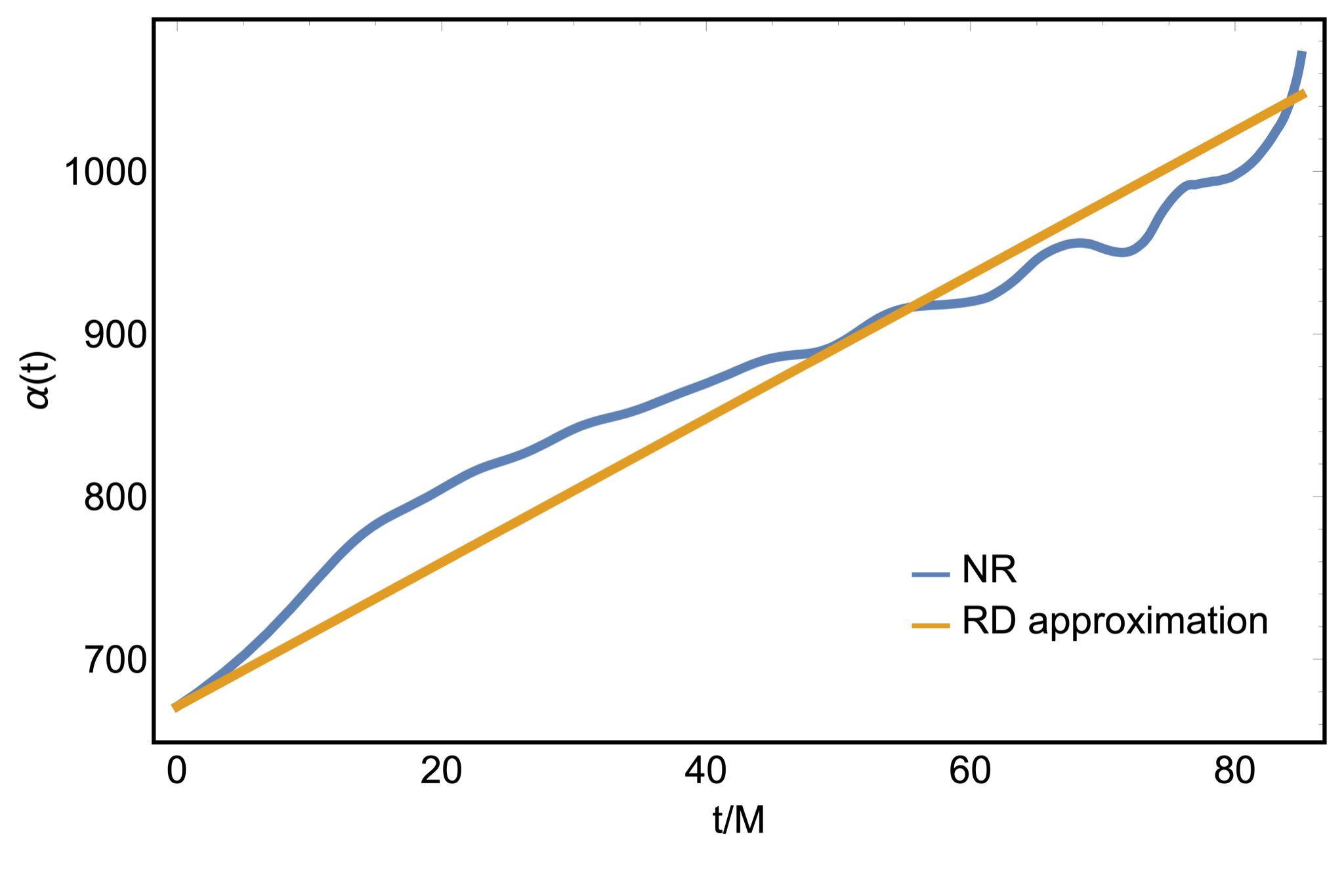}  
  \includegraphics[width=.85\linewidth]{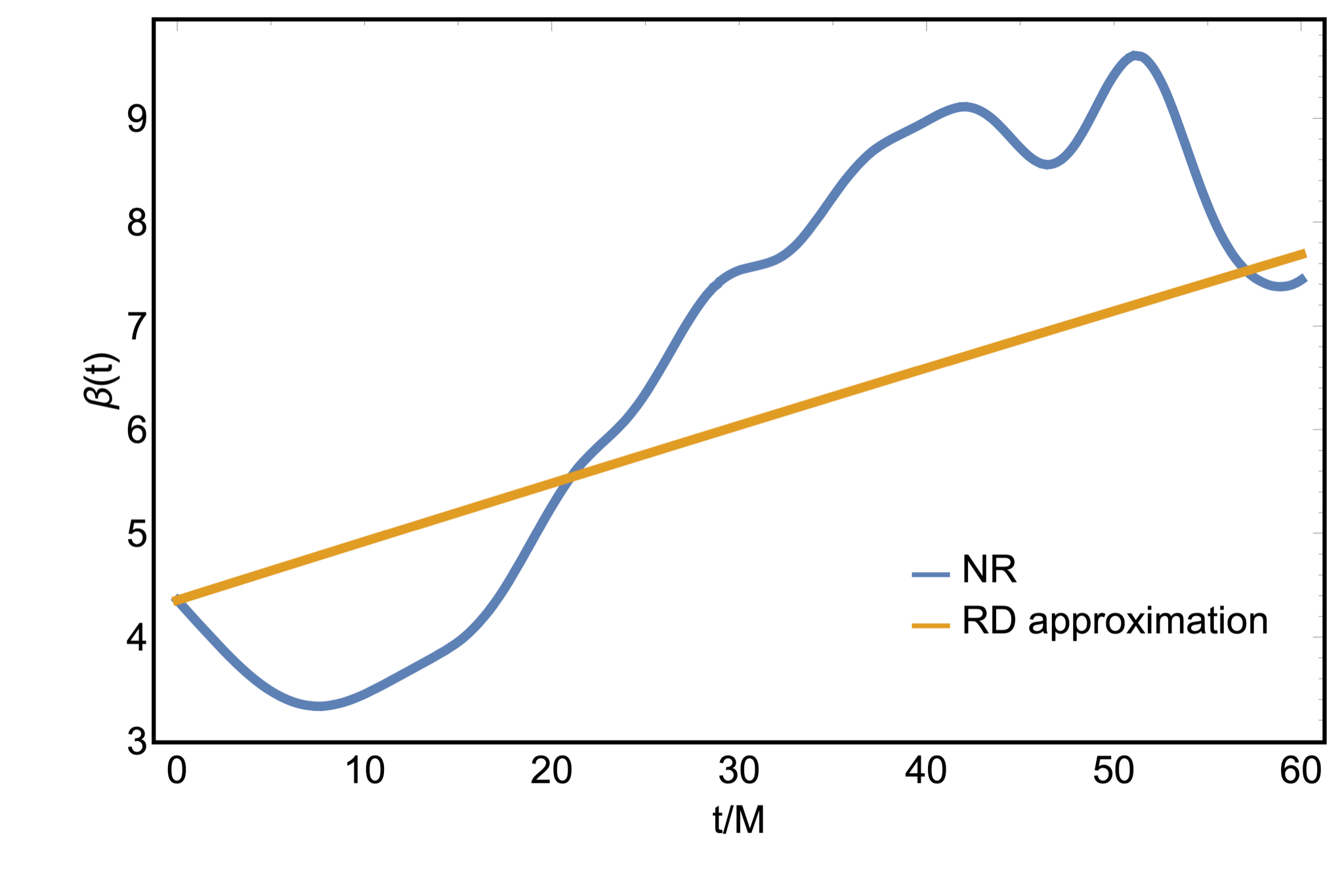}  
\caption{Comparison of the Euler angles during ringdown obtain from a precessing NR simulation (SXS:BBH:0015) and the analytical leading order approximation of eq. (\ref{eq:anglesrd}).}
\label{fig:anglesRD}
\end{figure}

\subsection{Polarisations construction}

In this section we outline the general steps for constructing the precessing waveform polarisations in a frame consistent with the LIGO Algorithm Library (LAL) conventions. For a more detailed explanation on the procedure, see \cite{phenomxphm}. Let us define the following reference frame systems: the coprecessing $\hat{\boldsymbol{\ell}}$ frame $\{\hat{\boldsymbol{X}},\hat{\boldsymbol{Y}},\hat{\boldsymbol{Z}}\}$, the $\hat{\boldsymbol{J}}$ frame $\{\hat{\boldsymbol{x}},\hat{\boldsymbol{y}},\hat{\boldsymbol{z}}\}$ and the detector wave frame $\{\hat{\boldsymbol{p}},\hat{\boldsymbol{q}},\hat{\boldsymbol{N}}\}$.
\subsubsection{$\hat{\boldsymbol{\ell}}$ frame}

The non-inertial $\hat{\boldsymbol{\ell}}$ frame is defined as having $\hat{\boldsymbol{Z}}$=$\hat{\boldsymbol{\ell}}$ and $\{\hat{\boldsymbol{X}},\hat{\boldsymbol{Y}}\}$ spanning the orbital plane, with $\hat{\boldsymbol{X}}$ in the direction from the heavier black hole to the lighter at some reference time. With the right hand rule convention, the third vector is defined as $\hat{\boldsymbol{Y}}=\hat{\boldsymbol{Z}}\times\hat{\boldsymbol{X}}$. A point $\hat{\boldsymbol{r}}$ on the unit sphere  will then have the corresponding spherical coordinates $\Theta=\arccos r_z$ and $\Phi=\arctan(r_y/r_x)$. In this system, the reference orbital phase of the system is 0 by definition. For defining the gravitational wave modes in this reference system, a basis of spin-weighted spherical harmonics \cite{swsh} has to be chosen. We choose the definition of \cite{Wiaux_2007}:
\begin{equation}
    _{-2}Y_{lm}(\Theta,\Phi)=S_{lm}(\Theta)e^{im\phi}
\end{equation}
so the spherical harmonic basis rotates counter-clockwise, respecting the right hand rule. With this definition, the positive $m$ SWSH modes $h_{lm}(t)$ rotate counter-clockwise (so $\dot{\psi}_{lm}<0$) and the negative modes rotate clockwise. Regarding the reference phase of the modes, a subtlety arises from historical reasons. In NR simulations, the same $\hat{\boldsymbol{\ell}}$ frame is employed, but the quantity obtained is the $\Psi_4$ scalar, the second time derivative of $h(t)$, and in this reference system the associated modes $\Psi_{4,lm}$ have reference wave phase equal to 0 by convention. However, the double time integration introduces an extra factor of $e^{i\pi}$ in $h$, i.e a global $\pi$ rotation of $h$, so the reference wave phase of the $h_{lm}$ modes in this frame and with this convention is $\psi_{\text{ref},lm}=\pi$ . The polarisations defined on a plane tangential to the sphere at a point $\hat{\boldsymbol{r}}=(\Theta,\Phi)$ in this frame will be:
\begin{equation}
    h(t;\Theta,\Phi)=h_{+}-i h_{\times}=\sum_l\sum_{m=-l}^{l}h_{lm}{}_{-2}Y^{\hat{\boldsymbol{\ell}}}_{lm}(\Theta,\Phi).
\end{equation}

\subsubsection{$\hat{\boldsymbol{J}}$ frame. Non-precessing case}

In the non-precessing limit, the $\hat{\boldsymbol{J}}$ frame is related to the $\hat{\boldsymbol{\ell}}$ frame by a rotation in the orbital plane. In the $\hat{\boldsymbol{J}}$ frame, $\hat{\boldsymbol{z}}=\hat{\boldsymbol{J}}=\hat{\boldsymbol{\ell}}$ but now the $\hat{\boldsymbol{x}}$ vector is defined as lying in the plane defined by $\hat{\boldsymbol{J}}$ and the line-of-sight vector between the Earth and the source, and $\hat{\boldsymbol{y}}=\hat{\boldsymbol{z}}\times\hat{\boldsymbol{x}}$ respecting again the right hand rule.
In this new reference frame, we can define a new basis of SWSH $_{-2}Y^{\hat{\boldsymbol{J}}}_{lm}(\theta,\phi)$.

Let be $\phi_{\text{ref}}$ the angle between $\hat{\boldsymbol{x}}$ and $\hat{\boldsymbol{X}}$ defined counter-clockwise from $\hat{\boldsymbol{x}}$ to $\hat{\boldsymbol{X}}$ and $\theta_{JN}$ the angle between $\hat{\boldsymbol{J}}$ and the line-of-sight. Then, the polarisations defined in the line-of-sight will be:
\begin{equation}
\begin{split}
    h_{+}-i h_{\times}&=\sum_l\sum_{m=-l}^{l}h_{lm}\ _{-2}Y^{\hat{\boldsymbol{\ell}}}_{lm}(\theta_{JN},-\phi_{\text{ref}})\\
    &=\sum_l\sum_{m=-l}^{l}e^{-im\phi_{\text{ref}}}h_{lm}\ _{-2}Y^{\hat{\boldsymbol{J}}}_{lm}(\theta_{JN},0).
\end{split}
\end{equation}

\subsubsection{$\hat{\boldsymbol{J}}$ frame. Precessing case}

In the precessing scenario, it is no longer true that $\hat{\boldsymbol{J}}=\hat{\boldsymbol{\ell}}$ for all times since $\hat{\boldsymbol{\ell}}$ is evolving. In fact, the time dependent relation between the two directions is given by the instantaneous Euler rotation described in Sec.\,\ref{sec:twistingup}. At the reference time, the direction $\hat{\boldsymbol{\ell}}$ in the $\hat{\boldsymbol{J}}$ frame is $\hat{\boldsymbol{\ell}}=(\cos\alpha_{\text{ref}}\sin\beta_{\text{ref}},\sin\alpha_{\text{ref}}\sin\beta_{\text{ref}},\cos\beta_{\text{ref}})$. For obtaining the polarisations, first we need to express the SWSH modes in the $\hat{\boldsymbol{J}}$ frame performing the rotation:
\begin{equation}
    h^{\hat{\boldsymbol{J}}}_{lm}=e^{im\alpha_{\text{ref}}}\sum_{m'}e^{im\gamma_{\text{ref}}}d^l_{mm'}(\beta_\text{ref})h^{\hat{\boldsymbol{\ell}}}_{lm'}.
\end{equation}
The polarisations are then given by:
\begin{equation}
\begin{split}
    h_{+}-i h_{\times}
    &=\sum_l\sum_{m=-l}^{l}h^{\hat{\boldsymbol{J}}}_{lm}\ _{-2}Y^{\hat{\boldsymbol{J}}}_{lm}(\theta_{JN},0).
\end{split}
\end{equation}
We see then that for recovering the non-precessing limit, where $\alpha$ is not well defined, we need to set $\alpha_{\text{ref}}=-\phi_{\text{ref}}$ and $\gamma_{\text{ref}}=0$.

\subsubsection{Detector frame}

These polarisations, which are transverse to the line-of-sight direction $\hat{\boldsymbol{N}}=(\sin\theta_{JN},0,\cos\theta_{JN})$, are defined in a plane tangential to the unit sphere at the intersection with the line-of-sight, with in plane reference system defined by the basis vectors $\hat{\boldsymbol{P}}=(\cos\phi,\sin\phi,0)$ and $\hat{\boldsymbol{Q}}=(-\sin\phi\cos\theta_{JN},\sin\phi\cos\theta_{JN},\sin\theta_{JN})$ expressed in the $\hat{\boldsymbol{J}}$ frame system. Then a transformation to the reference system defined in the detector's sky is needed to compute the polarisation content that the detector will see. 

For the detector wave frame, we employ the conventions in \cite{schmidt2017numerical} which define the frame in terms of the reference orbital angular momentum direction $\hat{\boldsymbol{L}}_{ref}$:
\begin{subequations}
\begin{align}
   &\hat{\boldsymbol{L}}_{ref}\cdot\hat{\boldsymbol{q}}=\sin\iota,\\ 
   &\hat{\boldsymbol{p}}=\frac{\hat{\boldsymbol{N}}\times\hat{\boldsymbol{q}}}{|\hat{\boldsymbol{N}}\times\hat{\boldsymbol{q}}|},
\end{align}
\end{subequations}
where $\iota=\arccos(\hat{\boldsymbol{L}}_{ref}\cdot\hat{\boldsymbol{N}})$ is the inclination between the reference orbital angular momentum direction and the line-of-sight. Denoting as $\xi$ the rotation angle in the polarisation plane between the $\{\hat{\boldsymbol{P}},\hat{\boldsymbol{Q}}\}$ and the $\{\hat{\boldsymbol{p}},\hat{\boldsymbol{q}}\}$ frames, the resulting polarisations in the detector frame are:
\begin{equation}
    \begin{pmatrix} h_{+}\\h_{\times}\end{pmatrix}_{\hat{\boldsymbol{N}}}=\begin{pmatrix} \cos\xi & \sin\xi\\-\sin \xi & \cos\xi \end{pmatrix}\begin{pmatrix} h_{+}\\h_{\times}\end{pmatrix}_{\hat{\boldsymbol{J}}}
\end{equation}



\section{\label{sec:sec6}Conclusions}

This work presents an IMR phenomenological model calibrated to NR simulations in time domain for the dominant modes $l=2$, $m=|2|$ of non-precessing BBH systems and its extension to precessing systems following the twisting up procedure. The underlying aligned-spin model combines an extension of the PN approximant TaylorT3 with higher pseudo-PN order terms calibrated to NR to describe the inspiral emission up to the MECO time, and phenomenological formulae for describing the merger-ringdown emission. The accuracy of the model, tested against a hybridized version of the calibration dataset with EOB and against the aligned-spin surrogate model NRSurHyb3dq8, is comparable to other state-of-the-art non-precessing models for the $l=2$, $m=|2|$ modes, IMRPhenomXAS and SEOBNRv4. The accuracy degrades in the early inspiral, due to the difficulty of extending TaylorT3 in a consistent way to improve the accuracy at high frequency, which neglect the parameter dependency of the ``merger'' time predicted by TaylorT3 and produces dephasing in the early inspiral. 

The precessing twisted up extension aims to provide flexibility in the description of the time dependent Euler rotation from the coprecessing to the J aligned precessing frame. It incorporates the effective single spin PN description at NNLO, the double spin MSA PN description and the numerical integration of the orbit-averaged spin evolution equations. It incorporates a flexible PN description of the norm of the orbital angular momentum up to 4PN with spin terms for the inspiral and a merger-ringdown description  from the numerical computation of the radiated angular momentum by the non-precessing $l=2$, $|m|=2$ modes. For the final state of the precessing waveform, it incorporates a simple geometrical computation of the final spin in terms of the non-precessing final spin and the in-plane spin components and it provides an approximate description of the Euler angles in the ringdown region.

The flexibility of the precessing extension will allow future systematic studies of the different PN approximations for the Euler angles and will offer a playground in which develop and test new features to describe more accurately the morphology of the signal from precessing BBH systems, for example the tracking of transitional precession, more accurate descriptions of the precessing final state, calibration in the NR regime of the Euler angles descriptions or the modelling of the equatorial asymmetry in precessing systems.

\section*{Acknowledgements}

We thank Geraint Pratten for discussions and carefully reading the manuscript and valuable feedback.
This work was supported by European Union FEDER funds, the Ministry of Science, 
Innovation and Universities and the Spanish Agencia Estatal de Investigación grants FPA2016-76821-P,     
RED2018-102661-T,    
RED2018-102573-E,    
FPA2017-90687-REDC,  
Vicepresid`encia i Conselleria d’Innovació, Recerca i Turisme, Conselleria d’Educació, i Universitats del Govern de les Illes Balears i Fons Social Europeu, 
Generalitat Valenciana (PROMETEO/2019/071),  
EU COST Actions CA18108, CA17137, CA16214, and CA16104, and
the Spanish Ministry of Education, Culture and Sport grants FPU15/03344 and FPU15/01319.
Marta Colleoni acknowledges funding from the European Union's Horizon 2020 research and innovation programme, under the Marie Skłodowska-Curie grant agreement No. 751492.
The authors thankfully acknowledge the computer resources at MareNostrum and the technical support provided by Barcelona Supercomputing Center (BSC) through Grants No. AECT-2019-2-0010, AECT-2019-1-0022, AECT-2019-1-0014, AECT-2018-3-0017, AECT-2018-2-0022, AECT-2018-1-0009, AECT-2017-3-0013, AECT-2017-2-0017, AECT-2017-1-0017, AECT-2016-3-0014, AECT2016-2-0009,  from the Red Española de Supercomputación (RES) and PRACE (Grant No. 2015133131). 
%
%

\begin{widetext}

\appendix

\section{Post-Newtonian quantities}

In this Appendix we present the post-Newtonian (PN) coefficients that appear in the expressions for the TaylorT3 wave frequency (\ref{eq:taylort3eq2}), the $l=2$, $m=2$ mode amplitude (\ref{eq:pnamp}) and the orbital angular momentum (\ref{eq:Lorb}), with $\delta m=\sqrt{1-4\eta}$, $m_1=(1+\delta m)/2$ and $m_2=(1-\delta m)/2$. 

\subsection{TaylorT3 coefficients}
\label{sec:appendixOmega}
In the adiabatic approximation post-Newtonian expressions for the orbital phase and the frequency can be obtained assuming energy conservation  \cite{PhysRevD.80.084043}
\begin{equation}
 \frac{d E}{dt} + \mathcal{F}=0,
\label{eq:eqt31}
\end{equation}
where $E$ represent the gravitational binding energy of the binary and $\mathcal{F}$ is the gravitational wave energy flux. In Eq. \eqref{eq:eqt31} we have neglected contributions coming from the flow of mass into the black holes as it enters at high post-Newtonian order. In the case of a non-precessing quasicircular binary the orbital energy and the flux can be expanded in terms of a PN expansion parameter describing the characteristic velocity of the binary
\begin{equation}
v=\left(M \frac{d \phi}{dt} \right)^{1/3},
\label{eq:eqt32}
\end{equation}
where $M$ is the total mass of the system and $\phi$ is the orbital phase. Given $\mathcal{F}(v)$  and $d E(v)/dt = dE(v)/dv (dv/dt)$ one can write 
\begin{equation}
\frac{dv}{dt}= -\frac{\mathcal{F}}{E'(v)}, 
\label{eq:eqt33}
\end{equation}
with $E'(v)\equiv dE/dv$. The freedom in expressing the right hand side of Eq. \eqref{eq:eqt33} as a perturbative series in the PN expansion parameter led to different PN approximants for the orbital phase \cite{PhysRevD.80.084043}. 
In this Appendix we extend the non-spinning version of the TaylorT3 approximant as reported in \cite{PhysRevD.80.084043} to include spin effects. The TaylorT3 approximant provides analytical expressions for the orbital phase and frequency as a function of time, 
\begin{equation}
\omega^{T3}(t)= \omega_0\sum^N_{k=1} \hat{\omega}_k, 
\label{eq:eqt34}
\end{equation}
where $\omega_0=\theta^3/(8 \pi M)$ and $\theta=\left[\eta (t_{\text{ref}-t})/(5M) \right]^{-1/8}$. In the TaylorT3 approximant $t_{\rm ref}$ is some reference time computed from the initial conditions of the binary ~\cite{PhysRevD.80.084043}. 

We introduce spin-orbit and spin-spin interactions up to $3.5$ PN order following the procedure of \cite{Brown:2007jx}, which includes those in the TaylorT2 approximant. Thus, in our case $N=7$ in Eq. \eqref{eq:eqt34} and the coefficients in the sum are given by the following expressions,
The following coefficients are a generalisation with spin-orbit and spin-spin interactions of the non-spinning expression that appears in \cite{PhysRevD.80.084043}.

\begin{subequations}
\begin{align}
\label{eq:TaylorT3coef}
\hat{\omega}_0 =& 1,\\
\hat{\omega}_1 =& 0,\\
\hat{\omega}_2 =& \frac{743}{2688}+\frac{11 \eta }{32},\\
\hat{\omega}_3 =& -\frac{3 \pi }{10}+\frac{113}{160} (m_ 1 \chi_1+m_ 2 \chi_2)-\frac{19}{80} \eta  (\chi_1+\chi_2),\\
\hat{\omega}_4 =& \frac{1855099}{14450688}-\frac{243 \left(m_ 1 \chi_1^2+m_ 2 \chi_2^2\right)}{1024}+\frac{56975 \eta }{258048}+\frac{3 \eta  \left(81 \chi_1^2-158 \chi_1 \chi_2+81 \chi_2^2\right)}{1024}+\frac{371 \eta ^2}{2048},\\
\hat{\omega}_5 =&-\frac{7729 \pi }{21504} + \frac{146597}{64512} (m_ 1 \chi_1+m_ 2 \chi_2) + \frac{13 \pi \eta}{256}-\frac{1213 \eta(\chi_1+\chi_2)}{1152}+\frac{7\eta}{128} \text{$\delta $m} (\chi_1-\chi_2)-\frac{17}{128} \eta ^2 (\chi_1+\chi_2),\\
\hat{\omega}_6 =&-\frac{720817631400877}{288412611379200}+\frac{107 \gamma_E }{280}+\frac{53 \pi ^2}{200}-\frac{6127 \pi  (m_ 1 \chi_1+m_ 2 \chi_2)}{6400}-\frac{16928263 \left(m_ 1 \chi_1^2+m_ 2 \chi_2^2\right)}{68812800}\nonumber\\
& \frac{25302017977}{4161798144}\eta-\frac{451 \pi ^2\eta}{2048}+\frac{1051 \pi \eta (\chi_1+\chi_2)}{3200}+\frac{23281001 \eta(\chi_1^2+\chi_2^2)}{68812800}-\frac{377345\eta \chi_1 \chi_2}{1376256}+\frac{453767\eta\delta  \left(\chi_1^2-\chi_2^2\right)}{4915200}\nonumber\\
& -\frac{30913\eta^2}{1835008}+\frac{335129\eta^2 \left(\chi_1^2+\chi_2^2\right)}{2457600}-\frac{488071\eta^2 \chi_1 \chi_2}{1228800}+\frac{107}{280} \log (2 \theta ),\\
\hat{\omega}_7 =&-\frac{188516689 \pi }{433520640}+\frac{6579635551 (m_ 1 \chi_1+m_ 2 \chi_2)}{650280960} +\frac{3663 \pi  \left(m_ 1 \chi_1^2+m_ 2 \chi_2^2\right)}{5120}-\frac{67493 \left(m_ 1 \chi_1^3+m_ 2 \chi_2^3\right)}{81920}\nonumber\\ 
&-\frac{97765 \pi  \eta }{258048}-\frac{1496368361 \eta(\chi_1+\chi_2)}{185794560} -\frac{3663 \pi\eta  \left(\chi_1^2+\chi_2^2\right)}{5120} + \frac{3537 \pi \eta \chi_1 \chi_2}{2560} + \frac{206917\eta \left(\chi_1^3+\chi_2^3\right)}{163840}\nonumber\\
&-\frac{192709\eta \chi_1 \chi_2 (m_1 \chi_1+m_2 \chi_2)}{81920} -\frac{28633921\eta \text{$\delta $m} (\chi_1-\chi_2)}{12386304} + \frac{71931\eta \text{$\delta $m} \left(\chi_1^3-\chi_2^3\right)}{163840} + \frac{141769\eta^2 \pi }{1290240}\nonumber\\
&-\frac{840149 \eta^2(\chi_1+\chi_2)}{15482880} -\frac{2219 \eta^2\left(\chi_1^3+\chi_2^3\right)}{40960} +\frac{1343\eta^2 \chi_1 \chi_2 (\chi_1+\chi_2)}{40960} + \frac{2591\eta^2 \text{$\delta $m} (\chi_1-\chi_2)}{46080}-\frac{12029 \eta^3(\chi_1+\chi_2)}{92160}
\end{align}
\end{subequations}

\subsection{Inspiral amplitude coefficients}
\label{sec:appendixAmp}
These are the PN coefficients entering in equation (\ref{eq:pnamp}) for the inspiral amplitude at 3.5 PN order:

\begin{subequations}
\begin{align}
\label{eq:inspampcoef}
\hat{h}_0 =& 1\\
\hat{h}_1 =& 0\\
\hat{h}_2 =& -\frac{107}{42}+\frac{55 \eta }{42}\\
\hat{h}_3 =& 2 \pi -\frac{2 (\chi_1+\chi_2)}{3} + \frac{2 \text{$\delta $m} (\chi_1-\chi_2)}{3 (m_ 1+m_ 2)}+\frac{2}{3}
   \eta  (\chi_1+\chi_2)\\
\hat{h}_4 =&-\frac{2173}{1512} -\frac{1069 \eta }{216} + \frac{2047 \eta ^2}{1512} + (m_1 \chi_1^2 + m_2 \chi_2^2) - \eta(\chi_1 - \chi_2)^2 \\
\hat{h}_5 =&-\frac{107 \pi }{21} + \frac{34 \pi  \eta }{21}-24 i \eta\\
\hat{h}_6 =&\frac{27027409}{646800} -\frac{856 \gamma_E }{105} +\frac{2 \pi^2}{3}+\frac{428 i \pi }{105}-\frac{428}{105} \log (16 x)-\frac{278185 \eta }{33264}+\frac{41 \pi ^2 \eta}{96}-\frac{20261 \eta ^2}{2772}+\frac{114635 \eta ^3}{99792}\\
\hat{h}_7 =& -\frac{2173 \pi }{756}-\frac{2495 \pi  \eta}{378}+\frac{14333 i \eta }{162} + \frac{40 \pi  \eta ^2}{27}-\frac{4066 i \eta ^2}{945}
\end{align}
\end{subequations}

\subsection{Orbital angular momentum coefficients}
\label{sec:appendixL}
These are the PN coefficients entering in equation (\ref{eq:Lorb}) for the PN description of the orbital angular momentum norm:

\begin{subequations}
\begin{align}
\label{eq:Lcoef}
l_0 =& 1\\
l_1 =& 0\\
l_2 =& \frac{3}{2}+\frac{\eta }{6}\\
l_3 =& \frac{5}{6}\eta(\chi_1 + \chi_2) - \frac{10}{3}(m_1\chi_1 - m_2\chi_2)\\
l_4 =& \frac{27}{8} - \frac{19 \eta }{8} + \frac{\eta ^2}{24} + m_1\chi_1^2 + m_2\chi_2^2 - \eta(\chi_1^2-\chi_2^2)\\
l_5 =& \frac{35}{8}\eta(\chi_1 + \chi_2)+\frac{7}{72}\eta(m_1\chi_1 + m_2\chi_2)-7(m_1\chi_1 + m_2\chi_2)\\
l_6 =& \frac{135}{16}-\frac{6889 \eta}{144} +\frac{41 \pi ^2 \eta }{24}+\frac{31 \eta ^2}{24}-\frac{7 \eta ^3}{1296}\\
l_7 =& -\frac{81}{4}(m_1\chi_1 + m_2\chi_2) + \frac{1119}{32}\eta(\chi_1 + \chi_2) + \frac{633}{32}\delta m\eta(\chi_1 - \chi_2) + \frac{3\eta}{2}(\chi_1 - \chi_2)(m_1\chi_1^2 + m_2\chi_2^2) \nonumber\\
& -\frac{43\eta^2}{4}(\chi_1 + \chi_2)-\frac{7\eta^2}{16}\delta m (\chi_1-\chi_2) - \frac{3\eta^2}{2}(\chi_1 + \chi_2)(\chi_1 - \chi_2)^2-\frac{\eta^3}{16}(\chi_1 + \chi_2)\\
l_8 =&\frac{2835}{128}+\frac{98869 \eta }{5760}-\frac{128\eta \gamma_E }{3}-\frac{6455 \pi ^2 \eta }{1536}+\frac{356035 \eta ^2}{3456}-\frac{2255 \pi ^2 \eta^2}{576} -\frac{215 \eta ^3}{1728}-\frac{55 \eta ^4}{31104}-\frac{64}{3} \eta  \log (16 v^2)
\end{align}
\end{subequations}

\end{widetext}

\bibliography{mybib}{}
\bibliographystyle{aipauth4-1}

\end{document}